\documentclass[conference]{IEEEtran}
\usepackage{cite}
\usepackage{amsmath,amssymb}
\usepackage{graphicx}
\usepackage{hyperref}
\usepackage{algorithm}
\usepackage{algpseudocode}
\usepackage{bbm}

\usepackage{amsmath, amssymb, amsfonts}
\usepackage{bm}          
\usepackage{mathtools}   
\usepackage{braket}      

\usepackage{graphicx}
\usepackage{subcaption}

\usepackage{booktabs}
\usepackage{multirow}
\usepackage{caption}

\usepackage{xcolor}
\definecolor{epistemic}{RGB}{31,119,180} 
\definecolor{aleatoric}{RGB}{255,127,14} 

\usepackage{xcolor}
\usepackage{colortbl}
\definecolor{lightgray}{gray}{0.95}

\usepackage{hyperref}

\usepackage{booktabs,tabularx,array} 

\begin{document}

\title{Trustworthy Quantum Machine Learning: A Roadmap for Reliability, Robustness, and Security in the NISQ Era}
\author{
\IEEEauthorblockN{Ferhat \"Ozgur Catak\IEEEauthorrefmark{1}, Jungwon Seo\IEEEauthorrefmark{1}, and Umit Cali\IEEEauthorrefmark{2}}
\IEEEauthorblockA{\IEEEauthorrefmark{1}Department of Electrical Engineering and Computer Science, University of Stavanger, Norway \\
Email: \{f.ozgur.catak, jungwon.seo\}@uis.no}
\IEEEauthorblockA{\IEEEauthorrefmark{2}School of Physics, Engineering and Technology, University of York, UK \\
Email: umit.cali@york.ac.uk}
}

\maketitle

\begin{abstract}
Quantum machine learning (QML) is as a promising paradigm for tackling computational problems that challenge classical AI. Yet, the inherent probabilistic behavior of quantum mechanics, device noise in NISQ hardware, and hybrid quantum--classical execution pipelines introduce new risks that prevent reliable deployment of QML in real-world, safety-critical settings. This research offers a broad roadmap for Trustworthy Quantum Machine Learning (TQML), integrating three foundational pillars of reliability: (i) uncertainty quantification for calibrated and risk-aware decision-making, (ii) adversarial robustness against classical and quantum-native threat models, and (iii) privacy preservation in distributed and delegated quantum learning scenarios. We formalize quantum-specific trust metrics grounded in quantum information theory, including a variance-based decomposition of predictive uncertainty, trace-distance-bounded robustness, and differential privacy for hybrid learning channels. To demonstrate feasibility on current NISQ devices, we validate a unified trust assessment pipeline on parameterized quantum classifiers, uncovering correlations between uncertainty and prediction risk, an asymmetry in attack vulnerability between classical and quantum state perturbations, and privacy–utility trade-offs driven by shot-noise and quantum channel noise. This roadmap seeks to define trustworthiness as a first-class design objective for quantum AI.
\end{abstract}


\section{Introduction}

The advent of noisy intermediate-scale quantum (NISQ) computing has catalyzed rapid progress in quantum machine learning (QML), demonstrating potential advantages in pattern recognition, optimization, and generative modeling \cite{biamonte2017quantum}. By harnessing quantum state representations and entanglement, QML could enable AI systems beyond the reach of purely classical computation. However, unlike classical deep learning where trust-enhancing practices such as calibration, robust learning, and privacy auditing are increasingly mature, QML remains at an early stage in incorporating trustworthiness as a fundamental design objective.

Trustworthiness poses special problems in quantum environments. First, the probabilistic nature of the Born rule introduces inherent uncertainty, requiring explicit separation of model error from quantum noise. Second, hybrid quantum--classical architectures create new adversarial surfaces involving input encoding, variational parameters, or even measurement operators. Third, data used for quantum learning is often sensitive (e.g., biomedical or financial), yet distributed settings restrict centralized acquisition and amplify risk of leakage through gradients or intermediate states. These factors necessitate a unified view of how reliability, security, and privacy should be achieved in QML.

This analysis delivers a thorough roadmap for Trustworthy QML (TQML) built upon three foundational pillars:
\begin{enumerate}
    \item \textbf{Uncertainty quantification (UQ)} to provide calibrated confidence estimates and risk-aware decision-making.
    \item \textbf{Adversarial robustness} to ensure resilience to malicious or noisy perturbations applied in classical or quantum domains.
    \item \textbf{Privacy preservation} to enable secure hybrid and federated quantum learning without exposing sensitive data.
\end{enumerate}

To support reproducibility and further research, we provide the full implementation and experimental configurations used in this study.\footnote{\url{https://github.com/ocatak/trustworthy-quantum-machine-learning}}

Trustworthiness has become essential for AI deployment in healthcare, autonomous systems, and cybersecurity \cite{amodei2016concrete,raji2020closing}, and the need is even more pressing in QML due to the interaction between quantum physics and ML decision pipelines. Establishing principled methods for quantifying and enforcing reliability, robustness, and privacy is therefore critical to enable responsible adoption of quantum AI.

Our long-term vision is to establish trustworthiness as a first-class requirement for QML—on par with accuracy and computational advantage. We advocate for quantum learning systems whose predictions can be explained, whose vulnerabilities can be quantified and mitigated, and whose use of sensitive data remains secure by design. By unifying uncertainty quantification, adversarial robustness, and privacy preservation under a common information-theoretic framework, we aim to guide the development of reliable, robust, and secure quantum AI throughout the NISQ era and beyond. Achieving this vision will require collaboration across quantum computing, machine learning, cybersecurity, and standardization communities to define shared metrics, benchmarks, and certification pathways for responsible quantum AI deployment.

\section{Pillar I: Adversarial Robustness in QML}
\label{sec:adv-qml}

Adversarial robustness has been extensively studied in classical ML~\cite{yan2018deep,xie2019feature,chen2020adversarial}, where small, carefully crafted perturbations to the input $x \in \mathbb{R}^d$ can cause a trained classifier $f_\theta$ to misclassify, even when perturbations are imperceptible to humans. This phenomenon is typically formalized through bounded perturbations $\|\delta\|_p \leq \epsilon$ and adversarial risk

\begin{equation}
    R_{\text{adv}}(f_\theta) \;=\; \mathbb{E}_{(x,y)\sim \mathcal{D}}\left[\max_{\|\delta\|_p \le \epsilon}\; \mathbbm{1}\{f_\theta(x+\delta)\neq y\}\right]
\end{equation}

In QML, the situation is analogous but richer due to the structure of Hilbert space, the role of quantum measurements, and inherent hardware noise \cite{gong2022universal}. Instead of perturbations in $\mathbb{R}^d$, we consider adversarial perturbations in quantum states, measurements, or parameters:
\begin{itemize}
    \item \textbf{Input-space analogy.} In classical AI, adversarial attacks modify $x \mapsto x+\delta$~\cite{szegedy2013intriguing,goodfellow2014explaining}. In QML, the adversary perturbs the quantum state $\rho_x \mapsto \rho'_x$ within a ball defined by a quantum distance measure (e.g., trace distance or infidelity). The classical $\ell_p$-norm corresponds to quantum divergences such as

    \begin{equation}
        D_{\text{tr}}(\rho,\rho') = \tfrac{1}{2}\|\rho - \rho'\|_1
    \end{equation}

    Thus, $\ell_p$-ball constraints map naturally to fidelity- or trace-distance balls in Hilbert space.
    
    \item \textbf{Model-space analogy.} In deep networks, parameter perturbations $\theta \mapsto \theta+\delta$ capture weight-space adversaries or poisoning. In QML, variational parameters of PQCs can be perturbed: $\theta \mapsto \theta+\delta$ with $\|\delta\|\leq \gamma$. Just as adversarial training in DNNs considers worst-case parameter gradients, quantum adversarial training considers gradients via parameter-shift rules.

    \item \textbf{Measurement-space extension.} Unlike classical models, QML introduces a new attack surface: the measurement process~\cite{saki2021qubit,debus2025entangled}. The positive operator-valued measure (POVM) elements $\{M_y\}$ can be adversarially altered, introducing a channel-level perturbation. This has no direct analogue in classical settings, but conceptually relates to adversarial sensor corruption.
\end{itemize}

\paragraph{Defenses: Parallels and Gaps.}
Many defense mechanisms from classical AI extend conceptually to QML, but require adaptation:
\begin{enumerate}
    \item \emph{Adversarial training.} In both cases, min–max optimization improves robustness~\cite{lu2020quantum}. Classically: $\min_\theta \mathbb{E}[\max_{\delta} \mathcal{L}(f_\theta(x+\delta),y)]$. Quantumly: $\min_\theta \mathbb{E}[\max_{D(\rho',\rho_x)\le \epsilon}\, \mathcal{L}(f_\theta(\rho'),y)]$.
    \item \emph{Regularization.} Weight decay, dropout, or Lipschitz control in DNNs have analogues in PQCs: parameter smoothing, Bayesian PQCs, or Lipschitz bounds via quantum Fisher information~\cite{wendlinger2024comparative,meyer2024robustness}.
    \item \emph{Randomization.} Input randomization or randomized smoothing in classical AI parallels randomized compiling in QML, which can suppress coherent adversarial effects~\cite{van2024techniques}.
\end{enumerate}

\paragraph{Robustness Metrics.}
In both classical and quantum domains, robustness is quantified by the accuracy drop under worst-case perturbations:

\begin{equation}
    \mathrm{RA}(\epsilon) \;=\; \Pr\big[\arg\max_y f_\theta(x+\delta)=y, \ \|\delta\|\leq\epsilon\big]
\end{equation}

for the classical computation, and

\begin{equation}
    \mathrm{RA}_\mathrm{Q}(\epsilon) \;=\; \Pr\big[\arg\max_y f_\theta(\rho')=y, \ D(\rho',\rho_x)\leq\epsilon\big]
\end{equation}

for the quantum computation. The formal structure is similar, but the perturbation geometry differs fundamentally.

\paragraph{Key Insight.}
The study of adversarial robustness in QML should therefore be seen as a continuation of classical robustness research, but situated in the probabilistic, noisy, and geometric setting of quantum information. The conceptual parallels provide a roadmap, while quantum-specific elements (state fidelity, measurement perturbations, hardware noise) highlight new frontiers that cannot be captured in classical models.

\begin{table*}[t]
\centering
\caption{Comparative Analysis of Adversarial Robustness Frameworks in Classical and Quantum Machine Learning}
\label{tab:adv-robustness}
\renewcommand{\arraystretch}{1.3}
\begin{tabular}{@{}lp{6cm}p{6.5cm}@{}}
\toprule
\textbf{Aspect} & \textbf{Classical ML / Deep Learning} & \textbf{Quantum Machine Learning} \\
\midrule
\addlinespace[2pt]

\multicolumn{3}{@{}l}{\textit{\textbf{Attack Formulations}}} \\
\addlinespace[2pt]

Perturbation Space 
& Input perturbations: $x \mapsto x+\delta$ with $\|\delta\|_p \leq \epsilon$ in $\mathbb{R}^d$ 
& Quantum state perturbations: $\rho \mapsto \rho'$ with $D_{\text{tr}}(\rho,\rho') \leq \epsilon$ or $F(\rho,\rho') \geq 1-\epsilon$ in Hilbert space $\mathcal{H}$ \\

\addlinespace[4pt]

Parameter Attacks 
& Weight poisoning: $\theta \mapsto \theta+\delta$ in neural networks; gradient manipulation in federated learning 
& Variational parameter shifts: $\theta \mapsto \theta+\delta$ in PQCs; circuit parameter poisoning in quantum federated learning \\

\addlinespace[4pt]

Unique Attack Surfaces 
& Limited to input space $\mathcal{X}$ and parameter space $\Theta$ 
& Additional vectors: POVM manipulation $\{M_y\} \mapsto \{M'_y\}$, hardware noise channel exploitation, measurement basis attacks \\

\addlinespace[4pt]

Adversarial Risk 
& $R_{\text{adv}} = \mathbb{E}_{(x,y)\sim\mathcal{D}}\left[\max_{\|\delta\|_p\leq\epsilon} \mathbb{I}\{f_\theta(x+\delta)\neq y\}\right]$ 
& $R_{\text{adv}}^{\text{Q}} = \mathbb{E}_{(x,y)\sim\mathcal{D}}\left[\max_{D(\rho',\rho_x)\leq\epsilon} \mathbb{I}\{f_\theta(\rho')\neq y\}\right]$ \\

\addlinespace[6pt]
\midrule
\addlinespace[2pt]

\multicolumn{3}{@{}l}{\textit{\textbf{Defense Mechanisms}}} \\
\addlinespace[2pt]

Training-Time Defenses 
& Adversarial training via min-max optimization; $\ell_2$ regularization; dropout; input data augmentation 
& Quantum adversarial training over trace-distance balls; parameter smoothing via Bayesian PQCs; variational noise adaptation \\

\addlinespace[4pt]

Inference-Time Defenses 
& Input preprocessing and randomization; ensemble predictions; certified defenses via randomized smoothing; Lipschitz bound enforcement 
& Randomized compiling; dynamical decoupling; Pauli twirling; quantum error mitigation; ensemble over circuit instances; Lipschitz bounds via quantum Fisher information \\

\addlinespace[6pt]
\midrule
\addlinespace[2pt]

\multicolumn{3}{@{}l}{\textit{\textbf{Evaluation Metrics}}} \\
\addlinespace[2pt]

Robustness Metrics 
& Robust accuracy: $\text{RA}(\epsilon) = \Pr[\arg\max_y f(x+\delta^*) = y]$; certified radius; attack success rate; PGD accuracy 
& Robust accuracy under quantum perturbations; shot-noise-aware robust accuracy; quantum certified radius; fidelity-based robustness margin \\

\addlinespace[4pt]

Calibration Metrics 
& Expected calibration error (ECE); reliability diagrams; Brier score decomposition 
& Quantum-aware ECE accounting for measurement uncertainty; shot-count-dependent calibration; entropy-based confidence under Born rule \\

\bottomrule
\end{tabular}
\vspace{2pt}
\begin{flushleft}
\footnotesize
\textit{Note}: $D_{\text{tr}}(\cdot,\cdot)$ denotes trace distance, $F(\cdot,\cdot)$ denotes quantum fidelity, PQC refers to parameterized quantum circuit, POVM denotes positive operator-valued measure, and QFI denotes quantum Fisher information.
\end{flushleft}
\end{table*}

The Lipschitz constant provides a quantitative measure of the smoothness of a model’s decision boundary. A classifier $f(x)$ is $K$-Lipschitz if changes in its output are bounded by $K$ times the perturbation in its input. Models with smaller Lipschitz constants are inherently more resistant to adversarial perturbations, since small input changes cannot induce large shifts in output predictions. This principle has been widely studied in classical deep learning, where Lipschitz constraints and regularization have been shown to improve adversarial robustness and calibration \cite{cisse2017parseval,2024arXiv241101644S}.

Adversarial robustness in QML can be understood as a natural extension of robustness research in classical AI. The conceptual foundation (e.g., worst-case perturbations, min–max training, and robustness–accuracy trade-offs) remains intact. However, QML introduces fundamentally new dimensions: quantum state geometry, measurement-channel vulnerabilities, and hardware-induced noise. These aspects not only broaden the adversarial landscape but also demand novel defenses grounded in quantum information theory. Thus, while classical robustness provides the roadmap, quantum adversarial robustness charts new terrain where physics, information theory, and ML intersect.

\section{Pillar II: Uncertainty Quantification in QML}
\label{sec:uq-qml}

Uncertainty quantification (UQ) forms a cornerstone of building trust in supervised QML \cite{guo2017calibration,low2014quantum,nguyen2020quantum,10.3389/fcomp.2021.662632,10321713,wendlinger2025old}. The probabilistic nature of quantum models originates from the \emph{Born rule}~\cite{nielsen2010quantum}, which states that the probability of obtaining a measurement outcome $y$ is given by the squared magnitude of the corresponding amplitude, i.e., $p(y)=|\langle y|\psi\rangle|^2$. Consequently, repeated executions (shots) of the same quantum circuit naturally produce stochastic outcome distributions. However, this intrinsic randomness alone does not fully characterize model reliability. Some variability reflects \emph{aleatoric uncertainty}~\cite{kendall2017uncertainties}, arising from irreducible sources such as overlapping class distributions and hardware-level imperfections (e.g., decoherence, gate errors, and state-preparation-and-measurement (SPAM) noise). Other variability is \emph{epistemic uncertainty}~\cite{kendall2017uncertainties}, which stems from incomplete knowledge of the parameterized quantum circuit (PQC), limited or biased training data, and underexplored regions of the feature Hilbert space. Finally, quantum models inherit \emph{technical uncertainty} from finite measurement sampling (\emph{shot noise}), which has no direct analogue in classical learning~\cite{cerezo2021variational}. 

Disentangling these categories is essential: aleatoric uncertainty defines the fundamental limits of predictability in noisy quantum devices, epistemic uncertainty highlights where models can be improved through better training or architectures, while technical uncertainty determines the variance floor imposed by quantum measurements. Effective UQ in QML therefore not only provides calibrated confidence estimates but also guides risk-aware decision-making, active data acquisition, and the design of more robust quantum–classical learning pipelines.

\subsection{Sources of Uncertainty}
\label{sec:uq-sources}

\begin{figure*}[htbp!]
  \centering
  \includegraphics[width=1.0\linewidth]{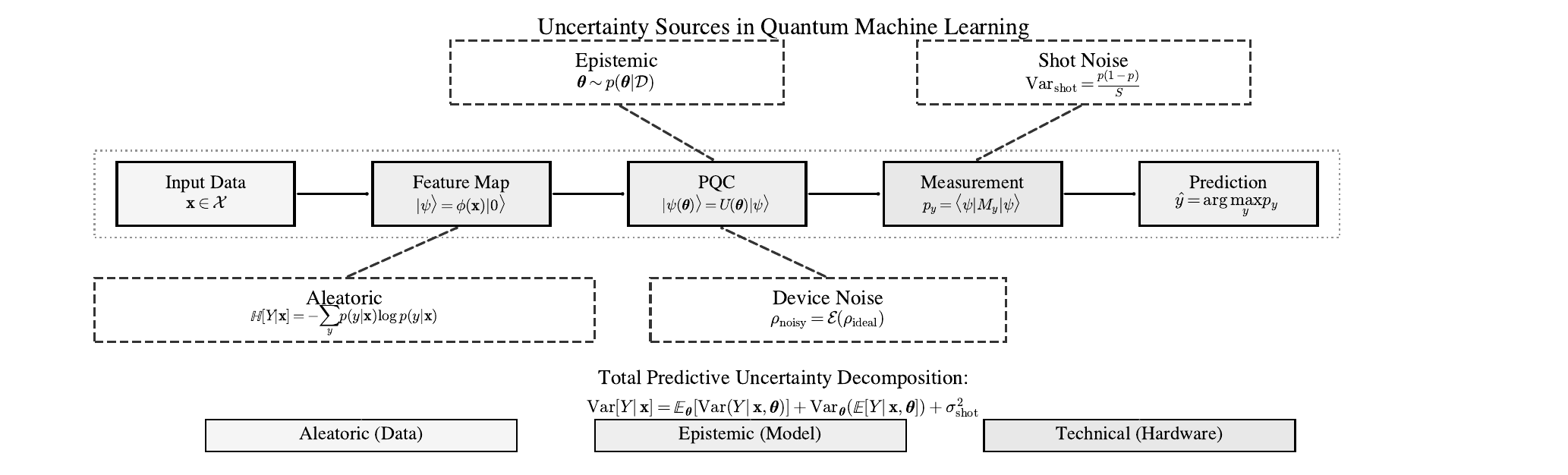}
  \caption{Uncertainty sources in a supervised QML pipeline. 
    Classical inputs are encoded into quantum states via a feature map, processed by a parameterized quantum circuit (PQC), and measured in a chosen basis before classical post-processing. 
    Predictive uncertainty arises from (i) \emph{aleatoric uncertainty} due to data ambiguity, (ii) \emph{epistemic uncertainty} from limited knowledge of PQC parameters, and (iii) \emph{technical uncertainty} due to device noise and finite sampling (shot noise).}
  \label{fig:pipeline}
\end{figure*}

Fig.~\ref{fig:pipeline} summarizes the supervised QML pipeline and highlights the fundamental sources of predictive uncertainty.  
Unlike in classical ML, where uncertainty is typically statistical or model-based, quantum models inherit additional variability from the physical principles governing their operation.  

\textbf{(i) Shot noise.}  
A direct consequence of the Born rule, shot noise arises when a PQC $U(\bm{\theta})$ is executed on an input state $\ket{\phi(x)}$ and measured in a basis $\{M_y\}$. The probability of observing outcome $y$ is
\begin{equation}
p_{\bm{\theta}}(y\!\mid\!x) \;=\; \langle \phi(x)|U^\dagger(\bm{\theta}) M_y U(\bm{\theta})|\phi(x)\rangle.
\end{equation}
In practice, $\hat{p}(y)$ is estimated from a finite number of shots $S$, introducing binomial variance
\begin{equation}
\mathrm{Var}_{\text{shot}}(y) \;=\; \frac{p(y)(1-p(y))}{S},
\end{equation}
which decreases with $S$ but never vanishes for finite samples.

\textbf{(ii) Device noise.}  
Real quantum processors deviate from ideal models due to decoherence, gate infidelities, crosstalk, and SPAM errors. Unlike shot noise, which is reducible with more samples, device noise represents irreducible stochasticity tied to hardware limitations, effectively broadening the predictive distribution.

\textbf{(iii) Parameter uncertainty.}  
Epistemic uncertainty arises because PQC parameters $\bm{\theta}$ are learned from limited and noisy data. Flat minima, barren plateaus, and restricted ansatz choices exacerbate indeterminacy~\cite{mcclean2018barren}. This form of uncertainty can be mitigated through richer datasets~\cite{gal2017deep}, better optimization~\cite{maddox2019simple}, Bayesian treatments~\cite{gal2016dropout}, or ensembles~\cite{lakshminarayanan2017simple}.

\textbf{(iv) Data-induced aleatoric uncertainty.}  
As in classical ML, data ambiguity (e.g., overlapping classes or noisy labels in $p(\mathbf{x},y)$)~\cite{kotelevskii2022nonparametric,northcutt2021confident} introduces irreducible unpredictability, independent of the quantum hardware.

Formally, the law of total variance decomposes predictive uncertainty:
\begin{equation}
\mathrm{Var}[Y\!\mid\!x] \;=\;
\underbrace{\mathbb{E}_{\bm{\theta}} \!\left[\mathrm{Var}(Y\!\mid\!x,\bm{\theta})\right]}_{\text{Aleatoric (data + device + shots)}}
\;+\;
\underbrace{\mathrm{Var}_{\bm{\theta}}\!\left(\mathbb{E}[Y\!\mid\!x,\bm{\theta}]\right)}_{\text{Epistemic (parameters)}}.
\label{eq:total-var}
\end{equation}

Figure~\ref{fig:pipeline} visually illustrates how these uncertainty sources propagate across the QML workflow.

\subsection{Uncertainty Quantification Methods in QML}

Several methods can be employed to estimate these uncertainties:
\begin{itemize}
    \item \textbf{Shot-based estimation:} Run circuits with different $S$ to quantify sampling variance.
    \item \textbf{Bayesian QML:} Treat PQC parameters as distributions $\bm{\theta}\sim p(\bm{\theta})$, using variational Bayes or stochastic-gradient Langevin dynamics~\cite{du2020expressive,nikoloska2024introduction}.
    \item \textbf{Quantum ensembles:} Train multiple PQCs with different initializations/ansatz to approximate epistemic uncertainty~\cite{schuld2018quantum,yang2023explainable}.
    \item \textbf{Information-theoretic metrics:} Predictive entropy 
    \(
    H(x)=-\sum_y \hat{p}(y|x)\log \hat{p}(y|x)
    \)
    captures total uncertainty, while mutual information decomposes epistemic vs. aleatoric components.
    \item \textbf{Credible intervals:} Bayesian posteriors (e.g., Dirichlet with Jeffreys prior) yield confidence bounds on $\hat{p}(y|x)$.
    \item \textbf{Quantum Fisher Information (QFI):} Local sensitivity to parameter perturbations indicates confidence calibration~\cite{yu2021experimental,ho2023stochastic}.
\end{itemize}

\begin{figure*}[!htbp]
  \centering
  \includegraphics[width=0.95\linewidth]{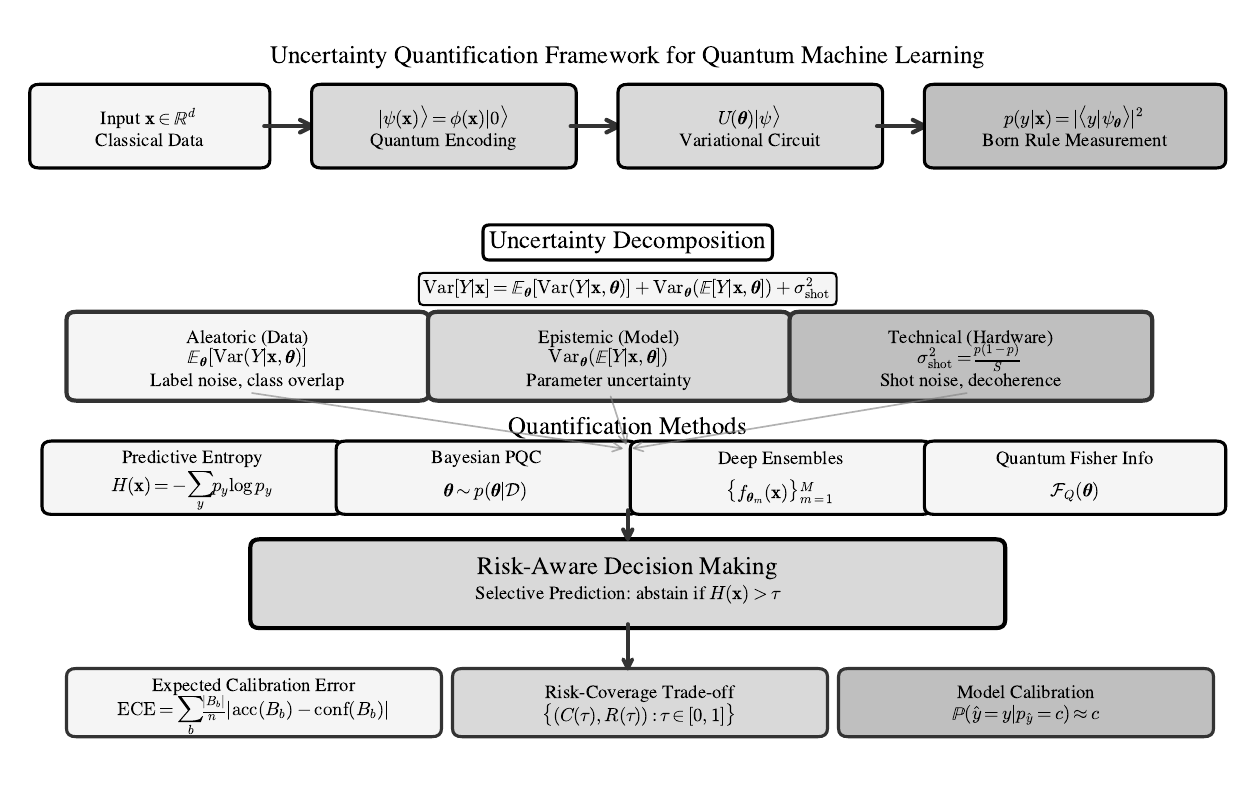}
  \caption{Uncertainty quantification methods in QML. 
  The diagram organizes predictive uncertainty into three primary sources: \emph{aleatoric} (data-related), \emph{epistemic} (model-related), and \emph{technical} (hardware-related). 
  Each source is linked to representative estimation strategies (e.g., shot-based sampling, Bayesian variational circuits, error mitigation), evaluation metrics (e.g., predictive entropy, mutual information, credible intervals), and downstream decision-making tools (e.g., selective prediction, active data acquisition, shot allocation). 
  The framework highlights how uncertainty analysis in QML is not only diagnostic but also prescriptive, shaping both model development and deployment.}
  \label{fig:uq-methods}
\end{figure*}

Figure~\ref{fig:uq-methods} illustrates the landscape of uncertainty quantification (UQ) methods in QML. 
As discussed in Section~\ref{sec:uq-sources}, uncertainty arises from three distinct sources: \emph{aleatoric} (data-level ambiguity such as label noise and class overlap), \emph{epistemic} (model-level uncertainty due to limited data, barren plateaus, or ansatz misspecification), and \emph{technical} (hardware-level stochasticity including shot noise, decoherence, and SPAM errors). 
The figure emphasizes that each source requires specialized estimation and calibration techniques. 

For \emph{aleatoric uncertainty}, predictive entropy, Brier score decomposition, and heteroscedastic likelihood modeling can quantify data-driven variability. 
For \emph{epistemic uncertainty}, Bayesian circuits, stochastic-gradient Langevin dynamics (SGLD), variational inference, and quantum ensembles offer principled methods to estimate model ignorance. 
For \emph{technical uncertainty}, shot-based variance estimation, multinomial confidence intervals, error mitigation protocols, and quantum Fisher information provide quantification tools tailored to quantum hardware. 

In particular, UQ in QML is not limited to diagnostic assessment: it enables \emph{decision support}. 
For instance, risk--coverage analysis allows models to abstain on uncertain predictions, active learning leverages epistemic uncertainty to guide data acquisition, and variance-aware shot allocation optimizes hardware usage under limited quantum resources. 
Thus, UQ methods play a dual role in QML—\emph{measuring reliability} and \emph{guiding trustworthy deployment}.

The practical implementation of our uncertainty quantification framework 
is formalized in Algorithm~\ref{alg:uq-quantum}. The procedure employs 
ensemble-based Monte Carlo sampling to separate epistemic uncertainty 
(arising from parameter indeterminacy) from aleatoric uncertainty 
(arising from inherent data ambiguity and shot noise). The computational 
complexity scales as $\mathcal{O}(M \cdot S \cdot d \cdot L)$ where 
$M$ is the ensemble size, $S$ is the shot count, $d$ is the input 
dimension, and $L$ is the circuit depth. 

\begin{algorithm}[!htbp]
\caption{Ensemble-Based Uncertainty Quantification for Variational Quantum Classifiers}
\label{alg:uq-quantum}
\begin{algorithmic}[1]
\Require Input $x \in \mathbb{R}^d$, shot budget $S \in \mathbb{N}^+$, ensemble size $M \in \mathbb{N}^+$
\Ensure Prediction $y^* \in \{1,\ldots,K\}$ with uncertainty decomposition

\State $\mathbf{P} \gets \mathbf{0}_{M \times K}$ \Comment{Initialize probability matrix}

\For{$m \gets 1$ \textbf{to} $M$}
    \State Sample initial parameters $\theta^{(m)} \sim \mathcal{N}(0, \sigma_{\text{init}}^2 \mathbf{I})$
    \State Prepare quantum state $\rho_x = \phi(x)\ket{0}\bra{0}\phi^\dagger(x)$
    \State Apply variational circuit $\rho_{\text{out}}^{(m)} = U(\theta^{(m)}) \rho_x U^\dagger(\theta^{(m)})$
    
    \For{$s \gets 1$ \textbf{to} $S$} \Comment{Shot-based sampling}
        \State Sample measurement outcome $y_s \sim \text{Born}(\rho_{\text{out}}^{(m)})$
        \State $n_y^{(m)} \gets n_y^{(m)} + \mathbbm{1}\{y_s = y\}$ for all $y$
    \EndFor
    
    \State $\hat{p}_y^{(m)} \gets n_y^{(m)} / S$ for all $y \in \{1,\ldots,K\}$
    \State $\mathbf{P}_{m,:} \gets [\hat{p}_1^{(m)}, \ldots, \hat{p}_K^{(m)}]$
\EndFor

\State $\bar{\mathbf{p}} \gets \frac{1}{M} \sum_{m=1}^M \mathbf{P}_{m,:}$ \Comment{Ensemble average}
\State $y^* \gets \arg\max_y \bar{p}_y$

\Statex
\State \textbf{Uncertainty Quantification:}
\State $H \gets -\sum_{y=1}^K \bar{p}_y \log \bar{p}_y$ \Comment{Total uncertainty (predictive entropy)}
\State $H_{\text{aleat}} \gets \frac{1}{M} \sum_{m=1}^M \left( -\sum_{y=1}^K \hat{p}_y^{(m)} \log \hat{p}_y^{(m)} \right)$ \Comment{Aleatoric}
\State $H_{\text{epist}} \gets H - H_{\text{aleat}}$ \Comment{Epistemic (mutual information)}

\Statex
\State \textbf{Credible Intervals:}
\State $\boldsymbol{\alpha} \gets [\sum_{m=1}^M n_y^{(m)} + 1]_{y=1}^K$ \Comment{Dirichlet posterior parameters}
\State Compute 95\% HPD credible regions from $\text{Dir}(\boldsymbol{\alpha})$

\State \Return $(y^*, H, H_{\text{aleat}}, H_{\text{epist}}, \text{credible intervals})$
\end{algorithmic}
\end{algorithm}

\subsection{Comparison with Classical AI}
The taxonomy of uncertainties in QML parallels that in classical AI but with crucial distinctions.  
Classical models experience \emph{aleatoric uncertainty} from noisy data and \emph{epistemic uncertainty} from limited training or model misspecification. \emph{Technical uncertainty} is typically negligible, arising only from numerical approximations. In contrast, QML inherits substantial technical uncertainty from the quantum hardware itself (shot noise, decoherence), making device calibration and error mitigation integral to reliable UQ. Thus, while Bayesian and ensemble methods transfer naturally to QML, uncertainty analysis in quantum pipelines must also explicitly account for hardware stochasticity.

\begin{table*}[t]
\centering
\caption{Comparison of uncertainty sources and quantification methods in classical AI and QML.}
\label{tab:uq-comparison}
\setlength{\tabcolsep}{7pt}
\renewcommand{\arraystretch}{1.25}
\begin{tabularx}{\textwidth}{
  @{} >{\raggedright\arraybackslash}p{2.4cm}
      >{\raggedright\arraybackslash}X
      >{\raggedright\arraybackslash}X
      >{\raggedright\arraybackslash}X @{}
}
\toprule
\textbf{Uncertainty Type} &
\textbf{Source} &
\textbf{Classical AI: common quantification} &
\textbf{QML: quantification / adaptation}
\\
\midrule
\textbf{Aleatoric} &
Data noise; label ambiguity; overlapping $p(\mathbf{x},y)$ &
Predictive entropy; Brier score and reliability/calibration curves; heteroscedastic likelihoods/regression &
As classical (entropy, calibration) \emph{plus} quantum-specific effects (SPAM/readout noise, hardware stochasticity); estimates combined with error-mitigation-aware calibration
\\[2pt]
\textbf{Epistemic} &
Limited data; model misspecification; parameter indeterminacy &
Bayesian neural networks; MC-dropout / deep ensembles; variational inference; mutual information &
Bayesian variational circuits; PQC ensembles; SGLD over $\boldsymbol{\theta}$; mutual-information decomposition between outputs and $\boldsymbol{\theta}$
\\[2pt]
\textbf{Technical} &
Numerical precision and optimization randomness (usually minor) &
Rarely treated explicitly; Monte-Carlo approximation error when applicable &
\emph{QML-specific:} finite-shot noise; decoherence; gate/crosstalk and SPAM errors; quantified via binomial/multinomial variance, Dirichlet posteriors for $\hat p(y\mid x)$; hardware error-mitigation (readout calibration, zero-noise extrapolation)
\\
\bottomrule
\end{tabularx}
\end{table*}

\subsection{Calibration and Reliability}
\label{sec:uq-calibration}

Beyond quantifying uncertainty, it is crucial to assess whether the reported confidence estimates are \emph{well calibrated}. A model is said to be calibrated if its predicted probabilities match the true empirical frequencies. For example, among all predictions where a classifier assigns $\hat{p}(y=1\mid \mathbf{x})=0.8$, approximately $80\%$ of the corresponding outcomes should indeed belong to class~1. Calibration is thus directly linked to model \emph{reliability}, i.e., the extent to which predicted confidence values can be trusted in decision-making.

\paragraph{Calibration Metrics.}  
Calibration can be measured using classical metrics adapted to QML outputs:
\begin{itemize}
    \item \textbf{Expected Calibration Error (ECE):} Partition the probability range into $B$ bins and compute
    \begin{equation}
    \mathrm{ECE} = \sum_{b=1}^B \frac{|S_b|}{n} \, \Big| \mathrm{acc}(S_b) - \mathrm{conf}(S_b) \Big|,
    \end{equation}
    where $\mathrm{acc}(S_b)$ and $\mathrm{conf}(S_b)$ denote the empirical accuracy and average predicted confidence in bin $b$.
    \item \textbf{Maximum Calibration Error (MCE):} The maximum absolute deviation across bins.
    \item \textbf{Brier Score:} The mean squared error between predicted probabilities and true one-hot labels.
\end{itemize}

\paragraph{Challenges in QML Calibration.}  
In QML, calibration is influenced not only by data and model design but also by quantum-specific uncertainties:
\begin{itemize}
    \item \emph{Shot noise} introduces additional variance in probability estimates, leading to under- or overconfidence depending on $S$.
    \item \emph{Device noise} can systematically bias predicted probabilities, degrading calibration even when epistemic uncertainty is low.
    \item \emph{Epistemic uncertainty} from limited training data can yield poorly calibrated models in underexplored regions of the Hilbert space.
\end{itemize}

\paragraph{Calibration Techniques.}  
Several methods can improve the reliability of QML confidence estimates:
\begin{itemize}
    \item \textbf{Classical post-hoc methods:} Temperature scaling and isotonic regression can be applied to QML predictive distributions $\hat{p}(y\mid \mathbf{x})$ after training.
    \item \textbf{Bayesian circuit approaches:} Treating PQC parameters as random variables provides better-calibrated posterior predictive distributions.
    \item \textbf{Ensemble averaging:} Combining outputs from multiple PQCs reduces variance and mitigates miscalibration due to local minima or barren plateaus.
    \item \textbf{Error mitigation:} Hardware-aware techniques (e.g., readout calibration, zero-noise extrapolation) directly reduce technical uncertainty, indirectly improving calibration.
\end{itemize}

\paragraph{Comparison with Classical AI.}  
In classical deep learning, calibration is often addressed through post-hoc scaling, Bayesian neural networks, or ensembles. These methods extend naturally to QML but must account for finite-shot stochasticity and hardware errors that are absent in classical settings. Consequently, calibration in QML is both a statistical and physical problem, requiring joint treatment of parameter uncertainty and quantum device reliability.

\medskip
In summary, calibration provides the bridge between \emph{uncertainty quantification} and \emph{trustworthy deployment}. Reliable QML systems must not only quantify uncertainty but also ensure that predictive probabilities faithfully reflect empirical correctness.

\subsection{Risk--Coverage Trade-offs}
\label{sec:uq-risk-coverage}

Uncertainty estimates enable QML models to trade off between \emph{risk} (expected error) and \emph{coverage} (fraction of inputs on which predictions are made) \cite{loquercio2020general}. Instead of always committing to a decision, the model may abstain on inputs with high predictive uncertainty. This abstention mechanism, common in selective prediction, is crucial for safety-critical applications of QML where erroneous decisions incur significant cost.

\paragraph{Formulation.}  
Given a prediction $\hat{y}(x)$ and confidence score $c(x)\in[0,1]$, a QML model can adopt a thresholding rule:
\begin{equation}
\hat{y}(x) =
\begin{cases}
\arg\max_y \hat{p}(y\mid x), & c(x) \geq \tau, \\
\text{abstain}, & c(x) < \tau,
\end{cases}
\end{equation}
where $\tau$ is a tunable confidence threshold. Increasing $\tau$ improves reliability (lower risk) but reduces the fraction of samples covered.

\paragraph{Risk--Coverage Curve.}  
Let $C(\tau)$ denote coverage, i.e., the fraction of inputs with $c(x)\geq\tau$, and $R(\tau)$ the empirical risk (e.g., classification error) on the retained set. Sweeping $\tau$ yields a \emph{risk--coverage curve}:

\begin{equation}
    \mathcal{R}\mathcal{C} = \{ (C(\tau), R(\tau)) \;:\; \tau \in [0,1] \}
\end{equation}

Well-calibrated models exhibit monotonically decreasing $R(\tau)$ with increasing $\tau$, indicating that uncertainty is a reliable proxy for risk.

\paragraph{Challenges in QML.}  
In quantum models, the risk--coverage analysis is complicated by additional stochasticity:
\begin{itemize}
    \item \textbf{Shot noise} may inflate predictive variance, leading to conservative abstention (lower coverage).
    \item \textbf{Device noise} can distort confidence estimates, degrading the alignment between uncertainty and true risk.
    \item \textbf{Epistemic uncertainty} from limited data may cause under-confidence in well-explored regions or over-confidence in underexplored ones.
\end{itemize}

\paragraph{Mitigation Strategies.}  
\begin{itemize}
    \item \emph{Ensembles and Bayesian circuits} improve the reliability of uncertainty estimates, yielding sharper risk--coverage curves.
    \item \emph{Post-hoc calibration} aligns predictive probabilities with empirical correctness, improving threshold selection.
    \item \emph{Error mitigation} reduces technical uncertainty, leading to more faithful coverage estimates.
\end{itemize}

\paragraph{Comparison with Classical AI.}  
In classical deep learning, risk--coverage analysis is widely used in selective classification, medical imaging, and autonomous driving. The same principle applies to QML, but the \emph{coverage frontier} is fundamentally influenced by quantum-specific uncertainties. In particular, while classical models suffer mainly from miscalibration, QML models must contend with both statistical and hardware-induced uncertainty, making error mitigation an essential part of improving the trade-off.

\medskip
In summary, the risk--coverage framework links uncertainty quantification to \emph{decision-making under uncertainty}. It provides a principled way to deploy QML models in real-world scenarios where safety and reliability are paramount.

\section{Pillar III: Privacy-Preserving QML}

Privacy is a cornerstone of trustworthy QML, particularly in domains such as healthcare, finance, and genomics where sensitive data cannot be freely centralized \cite{ren2025toward,watkins2023quantum}. Hybrid quantum--classical workflows introduce unique risks
of leakage through gradients, intermediate states, or measurement outcomes. To address these challenges, it is
crucial to define concrete threat models that guide the design of privacy-preserving techniques.

We identify four main threat models in QML:

\begin{itemize}
    \item \textbf{Centralized training risks:} Aggregating raw data into a single quantum--classical pipeline exposes sensitive information to compromise of the server or side-channel access to quantum measurements.

    \item \textbf{Distributed training risks:} In federated QML, gradients or parameter updates may leak private information about local datasets, enabling membership inference or reconstruction attacks similar to those in classical federated learning.

    \item \textbf{Hybrid communication risks:} Quantum models rely on both quantum and classical channels, making them vulnerable to man-in-the-middle interception or tomography-based leakage.

    \item \textbf{Delegated computation risks:} Offloading QML tasks to remote quantum servers requires trust in the provider. Without cryptographic safeguards, sensitive inputs and outputs may be exposed.
\end{itemize}

\textbf{Open Questions:}  
Several key challenges remain unresolved in privacy-preserving QML. First, defining \emph{quantum-native privacy guarantees} requires moving beyond classical differential privacy (DP), which is typically formalized as

\begin{equation}
    \Pr[\mathcal{M}(D) \in S] \leq e^{\varepsilon} \Pr[\mathcal{M}(D') \in S] + \delta,
\end{equation}

for neighboring datasets \(D, D'\) and mechanism \(\mathcal{M}\). Extending this notion to quantum states
\(\rho\) and \(\rho'\) under a quantum channel \(\mathcal{E}\) is non-trivial, as trace distance
\(\tfrac{1}{2}\|\mathcal{E}(\rho) - \mathcal{E}(\rho')\|_{1}\) or quantum relative entropy may serve as better divergence measures.

Additionally, there is a fundamental trade-off between \emph{accuracy and privacy} in quantum pipelines. In DP,
privacy budget \(\varepsilon\) controls the noise magnitude \(\mathcal{N}(0, \sigma^2)\) added to updates, while in QML,
device-induced noise \(\eta\) already perturbs outputs. The optimization problem can be written as

\begin{equation}
    \min_{\theta} \; \mathbb{E}_{(x,y) \sim D}\!\left[\mathcal{L}(f_{\theta}(x),y)\right]
+ \lambda \cdot \Phi(\varepsilon, \eta),
\end{equation}

where \(\mathcal{L}\) is the task loss and \(\Phi(\varepsilon,\eta)\) models the effective privacy–utility trade-off under
both cryptographic and physical noise. Characterizing \(\Phi\) in realistic hybrid quantum--classical settings is a
major open problem.

\begin{figure*}[t]
    \centering
    \includegraphics[width=\linewidth]{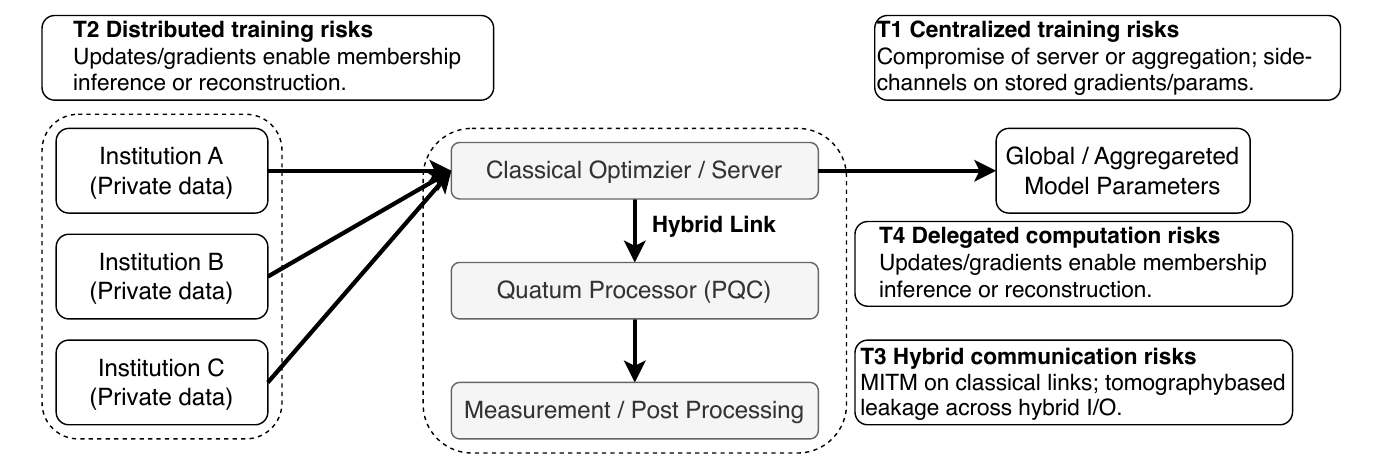}
    \caption{Privacy threat models in hybrid quantum--classical QML. 
    T1: centralized training risks (server or aggregation compromise); 
    T2: distributed training risks (updates or gradients leaking private data); 
    T3: hybrid communication risks (interception across quantum--classical channels); 
    T4: delegated computation risks (untrusted quantum providers).}
    \label{fig:ppqml-threats}
\end{figure*}

\subsection{Defense Strategies}

Potential strategies for privacy-preserving QML can be grouped into three complementary approaches: 
federated learning~\cite{mcmahan2017communication,chen2021federated}, homomorphic encryption~\cite{aono2017privacy,mahadev2020classical}, and differential privacy~\cite{wei2020federated,zhou2017differential}. Each provides a distinct protection layer, but also introduces unique challenges in the quantum setting.

\textbf{Federated Learning (FL):}  
In federated QML, each institution \(i \in \{1,\dots,N\}\) trains a local parameterized quantum circuit with 
parameters \(\theta_i\) on its private dataset \(D_i\). Instead of sharing raw data, participants communicate 
parameter updates \(\Delta \theta_i\) to an aggregator:

\begin{equation}
    \theta^{(t+1)} \;=\; \theta^{(t)} + \frac{1}{N}\sum_{i=1}^{N} \Delta \theta_i
\end{equation}

This preserves local data privacy, but gradients may still leak information~\cite{huang2021evaluating}. Adversaries could attempt to 
reconstruct samples \(x \in D_i\) by inverting \(\Delta \theta_i\). Secure aggregation protocols~\cite{mansouri2023sok}, often modeled as 
\(\mathrm{Enc}(\Delta \theta_i)\), ensure that only the aggregated update is revealed, reducing leakage risk. 
In the quantum setting, additional challenges arise due to the cost of repeated circuit evaluations and the 
need to coordinate hybrid quantum--classical resources.

\textbf{Homomorphic Encryption (HE):}  
Homomorphic encryption enables computation directly on encrypted data~\cite{cheon2017homomorphic}. Given a ciphertext 
\(c = \mathrm{Enc}(x)\), a homomorphic scheme allows:

\begin{equation}
\mathrm{Dec}(f(c)) \;=\; f(x)    
\end{equation}

where \(f(\cdot)\) represents the learning operation. In hybrid quantum--classical QML, this allows clients to 
send encrypted data or parameter updates to a remote quantum server without revealing sensitive information. 
Quantum homomorphic encryption (QHE)~\cite{tan2016quantum,zeuner2021experimental} further extends this idea to quantum states \(\rho\), where an 
encryption map \(\mathcal{E}\) ensures that

\begin{equation}
    \mathcal{D}\big( \mathcal{U}(\mathcal{E}(\rho)) \big) \;=\; \mathcal{U}(\rho)
\end{equation}

for a unitary \(\mathcal{U}\). While conceptually powerful, current QHE protocols are resource-intensive and 
practical only for limited circuits, leaving efficiency and scalability as open challenges.

\textbf{Differential Privacy (DP):}  
Differential privacy protects individual contributions by injecting calibrated noise into updates, measurement outcomes, or outputs~\cite{abadi2016deep}. In classical ML, this is formalized by \((\varepsilon,\delta)\)-DP:

\begin{equation}
    \Pr[\mathcal{M}(D) \in S] \;\leq\; e^{\varepsilon}\Pr[\mathcal{M}(D') \in S] + \delta
\end{equation}

for neighboring datasets \(D,D'\). In QML, datasets correspond to encoded quantum states 
\(\rho, \rho'\), and DP requires bounding

\begin{equation}
    \frac{1}{2}\big\|\mathcal{E}(\rho) - \mathcal{E}(\rho')\big\|_{1}
\end{equation}

where \(\mathcal{E}\) is the QML mechanism and \(\|\cdot\|_{1}\) is the trace norm. Designing 
quantum-native DP mechanisms is difficult because physical noise \(\eta\) from decoherence or finite sampling 
already perturbs outputs, making it hard to disentangle intentional DP noise from hardware imperfections~\cite{song2025towards}. 
Nevertheless, careful noise calibration may turn quantum stochasticity itself into a privacy resource.

To illustrate how these concepts can be combined in practice, Algorithm~\ref{alg:fedqml-dp} outlines a 
federated QML training loop with differential privacy noise injection. Each client trains a parameterized 
quantum circuit locally, perturbs its gradient updates with Gaussian noise, and shares only the noised updates 
with the server. The server then aggregates the contributions, ensuring that the global model benefits from 
distributed data without exposing sensitive local information.

\begin{algorithm}[t]
\caption{Federated QML with Differential Privacy Noise Injection}
\label{alg:fedqml-dp}
\begin{algorithmic}[1]
\Require $N$ clients with datasets $\{D_i\}_{i=1}^N$, initial parameters $\theta^{(0)}$, 
privacy budget $(\varepsilon, \delta)$, noise scale $\sigma$, number of rounds $T$
\For{$t = 0$ to $T-1$}
    \ForAll{client $i \in \{1,\dots,N\}$ in parallel}
        \State Encode local data $x \in D_i$ into quantum state $\rho_x$
        \State Execute PQC $U(\theta^{(t)})$ on $\rho_x$ and obtain measurement outcomes
        \State Compute local gradient update $\Delta \theta_i$
        \State Add Gaussian noise: 
        $\tilde{\Delta \theta}_i \gets \Delta \theta_i + \mathcal{N}(0, \sigma^2 I)$
        \State Send $\tilde{\Delta \theta}_i$ to server
    \EndFor
    \State Server aggregates updates: 
    $\theta^{(t+1)} \gets \theta^{(t)} + \frac{1}{N} \sum_{i=1}^N \tilde{\Delta \theta}_i$
\EndFor
\State \textbf{Output:} Final parameters $\theta^{(T)}$ with $(\varepsilon,\delta)$-DP guarantees
\end{algorithmic}
\end{algorithm}

\subsection{Open Research Directions}

\begin{itemize}
    \item Scalable FL protocols that balance communication efficiency and quantum resource constraints.
    \item Practical HE integration into variational training loops for secure outsourced computation.
    \item Quantum-specific DP formulations that distinguish between hardware noise and privacy-preserving noise.
    \item Standardized benchmarks and metrics for evaluating privacy in hybrid quantum--classical workflows.
\end{itemize}

\section{Experimental Demonstrations}\label{sec:demonstrations}

\subsection{Uncertainty Quantification in QML}\label{sec:demo_uncertainty}

To validate the proposed uncertainty quantification framework, we conduct detailed experiments on a binary classification task using a variational quantum classifier. We demonstrate how uncertainty metrics correlate with prediction correctness, how they evolve with shot count, and how spatial uncertainty patterns can guide model interpretation and deployment decisions.

\subsubsection{Experimental Setup}

\textbf{Dataset and Task.} We employ the two-moons dataset, a synthetic binary classification benchmark with inherent class overlap near the decision boundary. The dataset comprises 1500 samples split into training (60\%) and test (40\%) sets, with features standardized to zero mean and unit variance. This dataset is well-suited for uncertainty analysis due to its non-linear decision boundary and regions of high aleatoric uncertainty where classes overlap.

\textbf{Quantum Circuit Architecture.} Our parameterized quantum circuit (PQC) operates on 2 qubits with the following structure:
\begin{itemize}
    \item \textit{Feature encoding}: Angle encoding via RY rotations, mapping classical features $\mathbf{x} = (x_1, x_2)$ to quantum states.
    \item \textit{Variational layers}: A trainable ansatz with 4 variational parameters $\boldsymbol{\theta} = (\theta_1, \theta_2, \theta_3, \theta_4)$, comprising RY rotations and CNOT entangling gates.
    \item \textit{Measurement}: Pauli-Z expectation value on qubit 0 serves as the decision function, with binary predictions $\hat{y} = \text{sign}(\langle Z_0 \rangle)$.
\end{itemize}

\textbf{Training Protocol.} The PQC is trained using the Simultaneous Perturbation Stochastic Approximation (SPSA) optimizer~\cite{119632} with a squared hinge loss function for 50 iterations. The trained model achieves a test accuracy of 87.5\% on 600 test samples (525 correct predictions, 75 incorrect predictions), providing a substantial set of both correct and incorrect predictions for uncertainty analysis (Table~\ref{tab:classification_summary}).

\begin{table}[t]
\centering
\caption{Classification Performance Summary for Variational Quantum Classifier on Two-Moons Benchmark}
\label{tab:classification_summary}
\begin{tabular}{@{}lc@{}}
\toprule
\textbf{Metric} & \textbf{Value} \\
\midrule
\textit{Dataset Statistics} \\
\quad Total Test Samples & 600 \\
\quad Training Samples & 900 \\
\quad Features & 2 \\
\quad Classes & 2 \\
\addlinespace[4pt]
\textit{Classification Performance} \\
\quad Correct Predictions & 525 \\
\quad Incorrect Predictions & 75 \\
\quad Test Accuracy & 87.50\% \\
\addlinespace[4pt]
\textit{Quantum Circuit Parameters} \\
\quad Number of Qubits & 2 \\
\quad Circuit Depth & 4 \\
\quad Trainable Parameters & 4 \\
\quad Shot Count & 200 \\
\bottomrule
\end{tabular}
\vspace{2pt}
\end{table}

\textbf{Uncertainty Metrics.} We calculate multiple uncertainty quantification metrics from measurement outcomes:
\begin{itemize}
    \item \textit{Predictive entropy}: $H(\mathbf{x}) = -\sum_{y} \hat{p}(y|\mathbf{x}) \log \hat{p}(y|\mathbf{x})$ captures total uncertainty
    \item \textit{Variation ratio}: $\text{VR}(\mathbf{x}) = 1 - \frac{\max_y \#\{\hat{y}_s = y\}}{S}$ measures prediction disagreement
    \item \textit{Standard deviation}: $\sigma(\mathbf{x}) = \sqrt{\text{Var}[\hat{y}]}$ quantifies prediction variability
    \item \textit{Max confidence}: $\max_y \hat{p}(y|\mathbf{x})$ reflects model certainty
\end{itemize}
where $\hat{p}(y|\mathbf{x})$ is estimated from $S$ measurement shots, and each metric is computed over 200 shots per test sample.

\subsubsection{Statistical Validation of Uncertainty Estimates}

Table~\ref{tab:uncertainty_comparison} presents a detailed statistical comparison of uncertainty metrics between correct and incorrect predictions. The results strongly validate our uncertainty quantification framework with highly significant differences across all metrics.

\begin{table*}[!htbp]
\centering
\caption{Comparison of uncertainty metrics between correct and incorrect predictions. Values shown as mean $\pm$ standard deviation. $\Delta$ represents the difference (Incorrect - Correct). Statistical significance assessed using two-sample $t$-test.}
\label{tab:uncertainty_comparison}
\begin{tabular}{lccccc}
\toprule
Metric & Correct Pred. & Incorrect Pred. & $\Delta$ & Cohen's $d$ & $p$-value \\
\midrule
Predictive Entropy & 0.471 $\pm$ 0.316 & 0.900 $\pm$ 0.138 & 0.429 & 1.434 & 0.000 \\
Variation Ratio & 0.139 $\pm$ 0.137 & 0.359 $\pm$ 0.111 & 0.220 & 1.645 & 0.000 \\
Standard Deviation & 0.568 $\pm$ 0.285 & 0.928 $\pm$ 0.103 & 0.360 & 1.340 & 0.000 \\
Max Confidence & 0.861 $\pm$ 0.137 & 0.641 $\pm$ 0.111 & -0.220 & -1.645 & 0.000 \\
\bottomrule
\end{tabular}
\end{table*}

\textbf{Key Statistical Findings:}
\begin{enumerate}
    \item \textbf{Strong discriminative power}: All uncertainty metrics exhibit highly significant differences ($p < 0.001$) between correct and incorrect predictions, with large effect sizes. Predictive entropy shows a difference of $\Delta = 0.429$ (Cohen's $d = 1.434$), indicating that misclassified samples have nearly double the entropy of correct predictions ($0.900$ vs.\ $0.471$).
    
    \item \textbf{Consistent directionality}: Uncertainty metrics (entropy, variation ratio, standard deviation) are systematically higher for incorrect predictions, while confidence is systematically lower. This consistency across multiple metrics strengthens the reliability of uncertainty estimates.
    
    \item \textbf{Large effect sizes}: Cohen's $d$ values exceed 1.3 for all metrics, indicating that the separation between correct and incorrect predictions is substantial, well beyond the threshold for practical utility \cite{cohen2013statistical} (typically $d > 0.8$ is considered large).
    
    \item \textbf{Practical implications}: The strong correlation between uncertainty and correctness enables selective prediction strategies. For instance, abstaining on predictions with entropy $H(\mathbf{x}) > 0.7$ would reject most errors (high recall of incorrect predictions) while retaining the majority of correct predictions (high coverage).
\end{enumerate}

The breakdown in Table~\ref{tab:distribution_analysis} further reveals that incorrect predictions exhibit narrower standard deviations ($\sigma_{\text{incorrect}} = 0.138$ for entropy) compared to correct predictions ($\sigma_{\text{correct}} = 0.316$). Misclassifications cluster in high-uncertainty regions with relatively consistent uncertainty levels, while correct predictions span a broader uncertainty range.

\begin{table*}[!htbp]
\centering
\caption{Statistical comparison of uncertainty metrics between correct and incorrect predictions.}
\label{tab:distribution_analysis}
\begin{tabular}{lcccccc}
\toprule
Metric & Correct Mean & Correct Std & Incorrect Mean & Incorrect Std & t-statistic & p-value \\
\midrule
Predictive Entropy & 0.4712 & 0.3157 & 0.9003 & 0.1381 & 11.6709 & 0.0000 \\
Max Confidence & 0.8610 & 0.1370 & 0.6406 & 0.1112 & -13.3797 & 0.0000 \\
Variation Ratio & 0.1390 & 0.1370 & 0.3594 & 0.1112 & 13.3797 & 0.0000 \\
\bottomrule
\end{tabular}
\end{table*}

\begin{figure}[t]
    \centering
    \includegraphics[width=\linewidth]{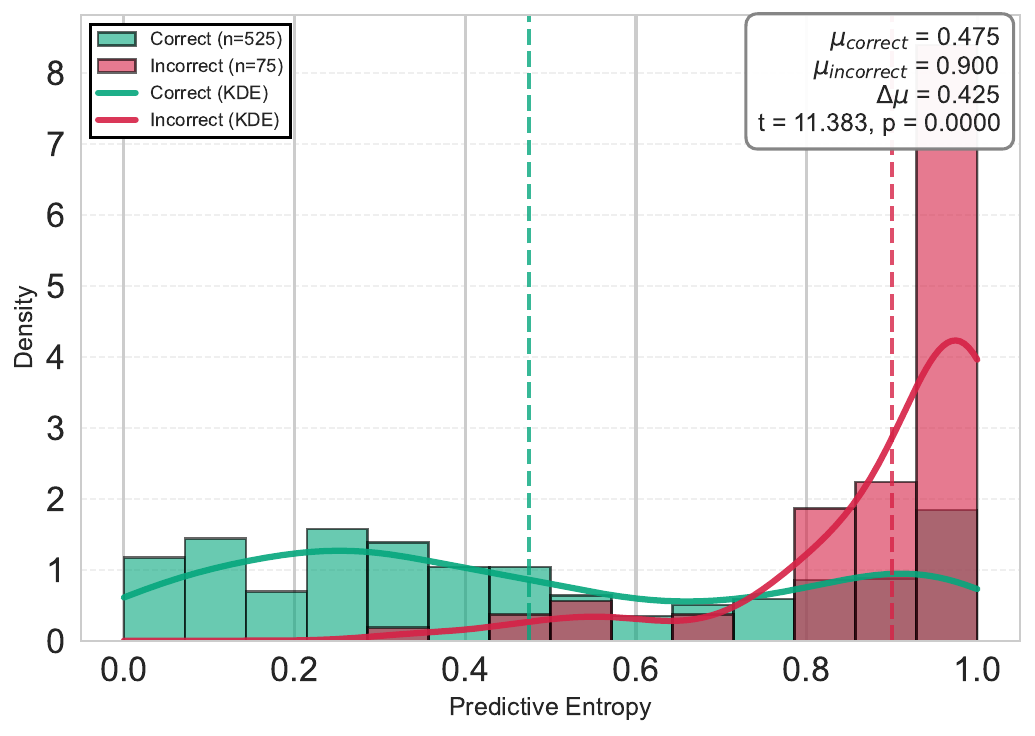}
    \caption{Distribution of predictive entropy for correct vs.\ incorrect predictions. Kernel density estimation (KDE) curves show clear bimodal separation: misclassified samples concentrate at high entropy ($\mu_{\text{incorrect}} = 0.900$), while correct predictions span a broader range with lower mean ($\mu_{\text{correct}} = 0.475$). The statistical test confirms this difference is highly significant ($t = 11.383, p < 0.0001$), validating entropy as a reliable proxy for prediction correctness. Dashed vertical lines indicate distribution means.}
    \label{fig:entropy_distribution}
\end{figure}

Figure~\ref{fig:entropy_distribution} presents the probability density functions of predictive entropy for correct versus incorrect predictions, estimated via kernel density estimation (KDE). The visualization reveals several critical insights into the distributional characteristics of uncertainty.

\textbf{Distributional Characteristics:}
\begin{itemize}
    \item \textbf{Bimodal separation}: The KDE curves show minimal overlap, with incorrect predictions forming a narrow, concentrated peak near maximum entropy ($H \approx 1.0$), while correct predictions exhibit a broader, left-skewed distribution concentrated at lower entropy values ($H < 0.6$). The t-statistic of 11.383 confirms this separation is far beyond random chance.
    
    \item \textbf{Threshold selection}: The clear separation suggests a natural decision threshold around $H \approx 0.7$ for selective prediction. Samples with entropy above this threshold are predominantly incorrect, enabling risk-averse deployment strategies that abstain on uncertain predictions.
    
    \item \textbf{Residual uncertainty}: The tail of correct predictions extending toward higher entropy values ($0.6 < H < 0.9$) represents inherent aleatoric uncertainty—samples near the decision boundary where even a well-calibrated model cannot achieve high confidence due to feature ambiguity and class overlap.
\end{itemize}

This distributional analysis validates a fundamental hypothesis of trustworthy QML: \textit{well-designed uncertainty metrics should exhibit strong separation between reliable and unreliable predictions}. The observed separation (Cohen's $d = 1.434$) significantly exceeds the threshold for practical decision-making applications in safety-critical domains.

\begin{figure*}[t]
    \centering
    \begin{subfigure}[b]{0.32\linewidth}
        \centering
        \includegraphics[width=\linewidth]{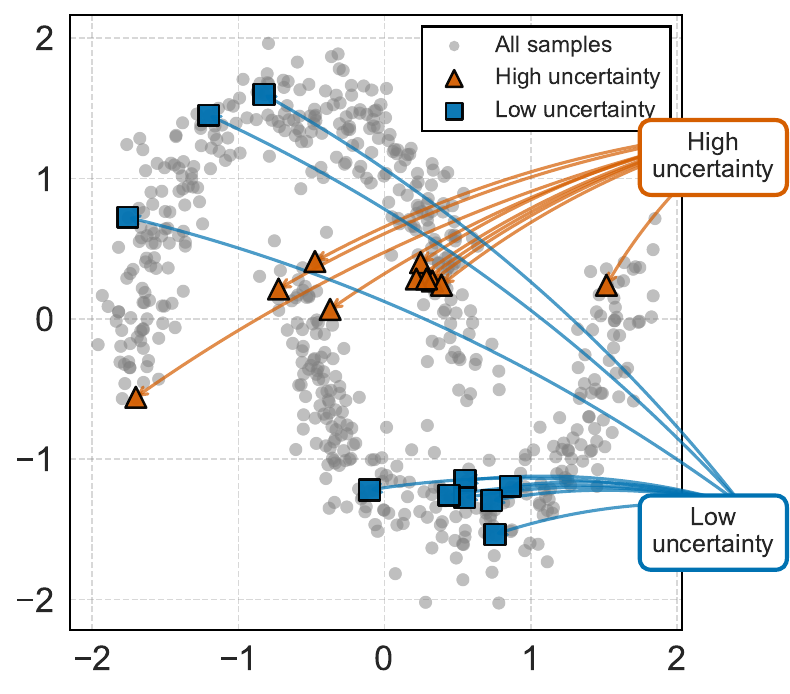}
        \caption{$n = 10$ shots}
        \label{fig:uncertainty_10shots}
    \end{subfigure}
    \hfill
    \begin{subfigure}[b]{0.32\linewidth}
        \centering
        \includegraphics[width=\linewidth]{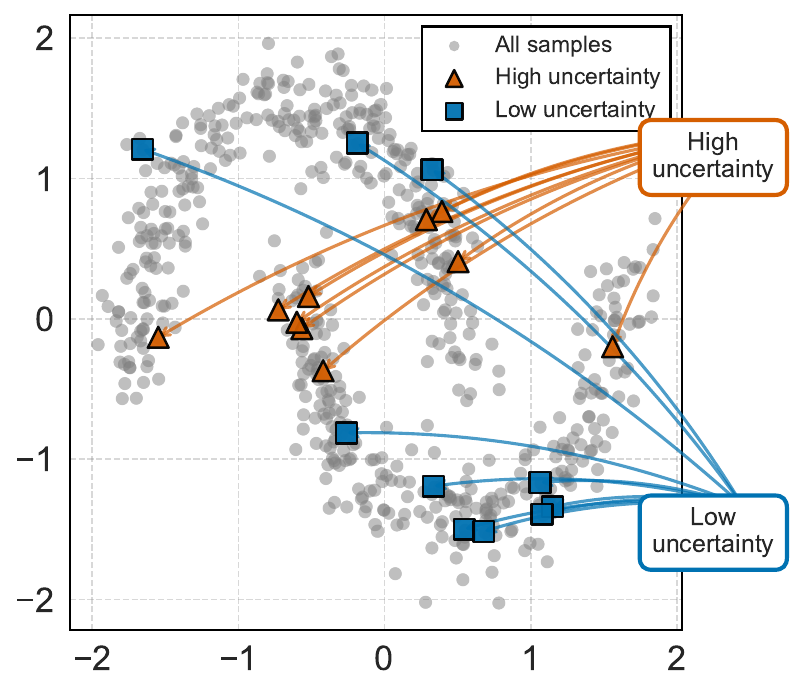}
        \caption{$n = 50$ shots}
        \label{fig:uncertainty_50shots}
    \end{subfigure}
    \hfill
    \begin{subfigure}[b]{0.32\linewidth}
        \centering
        \includegraphics[width=\linewidth]{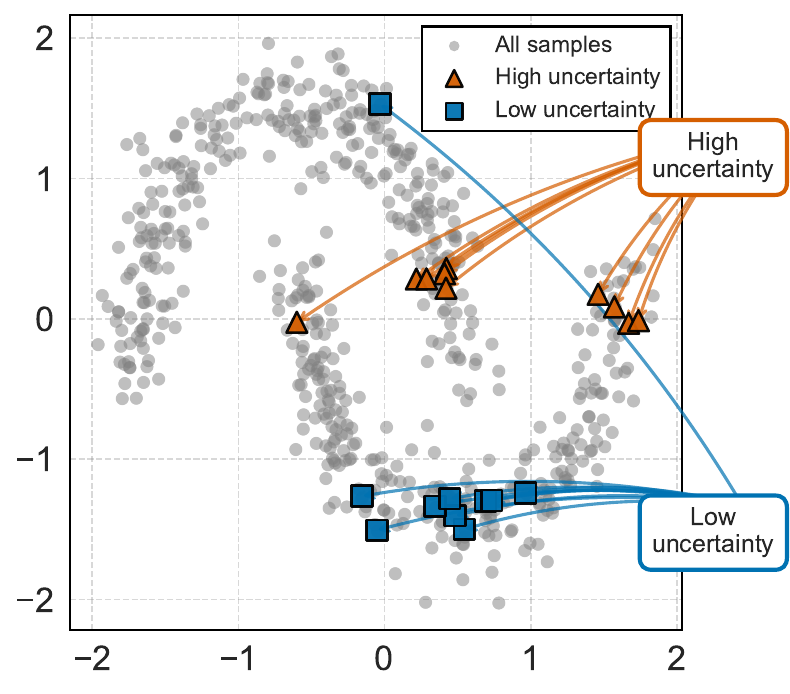}
        \caption{$n = 100$ shots}
        \label{fig:uncertainty_100shots}
    \end{subfigure}
    \hfill
    \begin{subfigure}[b]{0.32\linewidth}
        \centering
        \includegraphics[width=\linewidth]{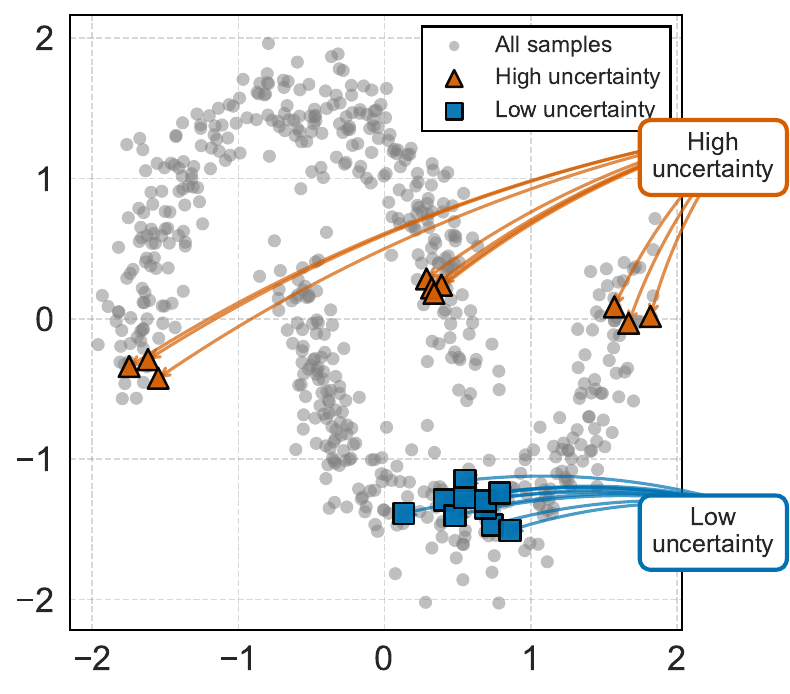}
        \caption{$n = 200$ shots}
        \label{fig:uncertainty_200shots}
    \end{subfigure}
    \hfill
    \begin{subfigure}[b]{0.32\linewidth}
        \centering
        \includegraphics[width=\linewidth]{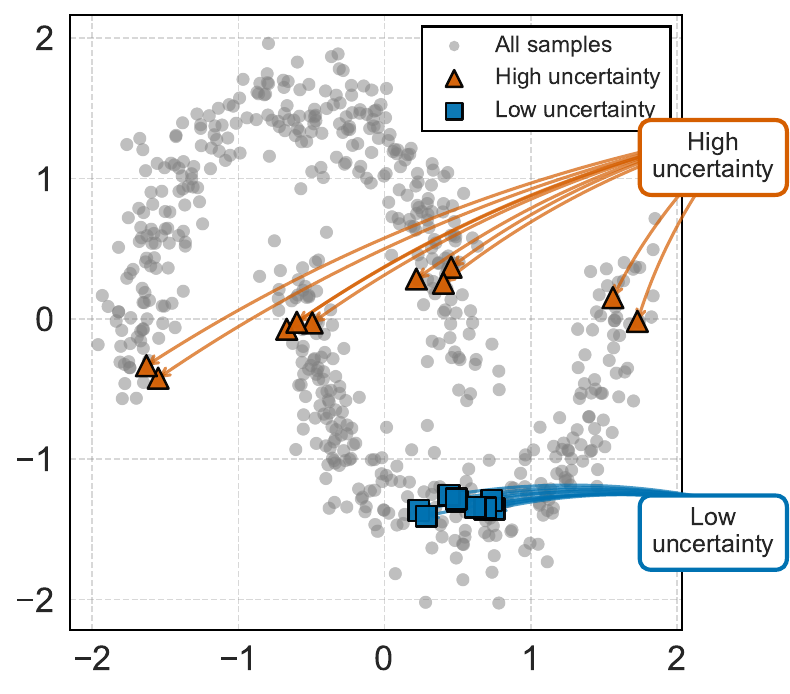}
        \caption{$n = 500$ shots}
        \label{fig:uncertainty_500shots}
    \end{subfigure}
    \hfill
    \begin{subfigure}[b]{0.32\linewidth}
        \centering
        \includegraphics[width=\linewidth]{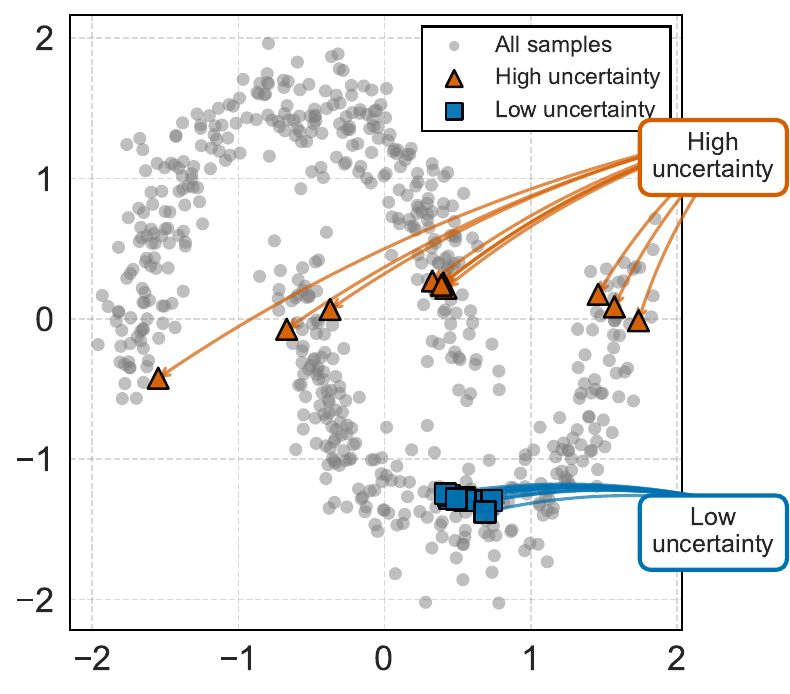}
        \caption{$n = 1000$ shots}
        \label{fig:uncertainty_1000shots}
    \end{subfigure}
    
    \caption{Spatial distribution of predictive uncertainty across feature space for varying shot counts. Orange triangles denote the 10 samples with the highest entropy (high uncertainty); blue squares indicate the 10 samples with the lowest entropy (high confidence). Arrows connect samples to annotation boxes. High-uncertainty samples consistently localize along the non-linear decision boundary, while confident predictions occur in class-dense regions. As shot count increases from 50 to 1000, the spatial pattern stabilizes, demonstrating a reduction in technical (shot noise) uncertainty while preserving the underlying epistemic and aleatoric structure.}
    \label{fig:uncertainty_spatial}
\end{figure*}

Figures~\ref{fig:uncertainty_10shots}--\ref{fig:uncertainty_1000shots} visualize how uncertainty localizes in the two-dimensional feature space across varying shot counts ($n \in \{10, 50, 100, 200, 500, 1000\}$). Each plot highlights the 10 samples with the highest predictive entropy (orange triangles) and 10 samples with the lowest entropy (blue squares), connected via arrows to annotation boxes.

\textbf{Spatial Analysis Findings:}
\begin{enumerate}
    \item \textbf{Decision boundary localization}: High-uncertainty samples (orange triangles) consistently cluster along the non-linear boundary separating the two crescent-shaped classes. This spatial pattern confirms that epistemic uncertainty concentrates where the model has limited discriminative information, a hallmark of well-calibrated uncertainty estimates. The boundary follows the region where classes overlap, exactly where we expect inherent ambiguity.
    
    \item \textbf{Confidence in class cores}: Low-uncertainty samples (blue squares) reside deep within class-specific regions, far from the decision boundary. These samples receive confident, stable predictions with $H(\mathbf{x}) < 0.2$ regardless of shot count, reflecting both low aleatoric uncertainty (clear class membership) and low epistemic uncertainty (model has learned this region well).
    
    \item \textbf{Shot-dependent stabilization}: At $n = 50$ shots (Figure~\ref{fig:uncertainty_50shots}), the identification of uncertain samples exhibits higher variability across repeated experiments due to shot noise—a manifestation of technical uncertainty. As shot count increases to $n = 1000$ (Figure~\ref{fig:uncertainty_1000shots}), the spatial pattern stabilizes significantly, with consistent identification of the same boundary-region samples as uncertain. This progressive stabilization quantifies the reduction in technical uncertainty with increased sampling, as predicted by the binomial variance formula:
    \begin{equation}
    \text{Var}_{\text{shot}}(y) = \frac{p(y)(1-p(y))}{n}.
    \end{equation}
    
    \item \textbf{Symmetric uncertainty structure}: The spatial distribution exhibits approximate symmetry across both crescent-shaped classes, indicating that the model does not exhibit systematic bias toward one class. Both class boundaries are associated with similar uncertainty magnitudes, suggesting balanced learning.
    
    \item \textbf{Actionable insights for deployment}: The clear spatial localization provides interpretable explanations for model behavior. In safety-critical applications, flagging boundary-region samples for manual review or abstention would substantially reduce error rates. Additionally, active learning strategies could prioritize label acquisition in these high-uncertainty regions to maximize information gain per query.
\end{enumerate}

Table~\ref{tab:statistical_summary} quantifies how accuracy and uncertainty metrics evolve as shot count increases from 50 to 1000 shots. The analysis reveals critical insights into the trade-offs between technical uncertainty and computational cost.

\begin{table*}[!htbp]
\centering
\caption{Statistical summary of metric evolution with shot count. Pearson correlation coefficients and p-values assess the strength of linear relationships.}
\label{tab:statistical_summary}
\begin{tabular}{lccccc}
\toprule
Metric & Initial Value & Final Value & Change (\%) & Correlation with Shots ($r$) & $p$-value \\
\midrule
Accuracy & 0.8733 & 0.8783 & 0.57 & 0.2215 & 0.0004 \\
Entropy & 0.5126 & 0.5281 & 3.02 & 0.5780 & 0.0000 \\
Variation Ratio & 0.1628 & 0.1680 & 3.17 & 0.3887 & 0.0000 \\
Confidence & 0.8372 & 0.8320 & $-$0.62 & $-$0.3887 & 0.0000 \\
\bottomrule
\end{tabular}
\end{table*}

\textbf{Shot Count Dynamics:}
\begin{itemize}
    \item \textbf{Modest accuracy gains}: Test accuracy improves slightly from 87.33\% to 87.83\% ($\Delta = 0.57\%$) as shots increase, with a weak positive correlation ($r = 0.22, p < 0.001$). This modest improvement suggests that technical (shot) noise is not the dominant source of classification errors; rather, errors stem primarily from epistemic and aleatoric uncertainties inherent to the task and model architecture.
    
    \item \textbf{Uncertainty metric shifts}: Predictive entropy increases by 3.02\% ($0.513 \to 0.528$) with strong positive correlation to shot count ($r = 0.58, p < 0.001$). This counterintuitive trend—increasing uncertainty with more shots, arises because higher shot counts provide more accurate estimates of the true probability distribution, revealing underlying uncertainty that low-shot estimates underestimate due to discretization effects. In other words, with fewer shots, the model appears artificially confident; more shots expose the true uncertainty.
    
    \item \textbf{Variation ratio consistency}: The variation ratio increases by 3.17\% with moderate correlation ($r = 0.39$), paralleling the entropy trend. This indicates that the model's predictions become more calibrated—not more confident—as sampling improves.
    
    \item \textbf{Confidence calibration}: Max confidence decreases by 0.62\% as shots increase, with negative correlation ($r = -0.39$). This aligns with the entropy increase: better sampling reveals the model is less certain than low-shot estimates suggested, improving calibration at the cost of apparent confidence.
\end{itemize}

\textbf{Practical Resource Allocation}: The correlation analysis reveals diminishing returns beyond $n \approx 500$ shots for this task. The accuracy improvement from 500 to 1000 shots is less than 0.1\%, while computational cost doubles. For resource-constrained quantum hardware, adaptive shot allocation strategies—allocating more shots to boundary-region samples and fewer to confident samples—could optimize the uncertainty-cost trade-off.

\subsubsection{Detailed Uncertainty Profile}

Table~\ref{tab:uncertainty_full} provides the complete statistical profile, including median values and all four uncertainty metrics, offering deeper insights into distributional characteristics.

\begin{table*}[t]
\centering
\caption{Detailed uncertainty metrics analysis comparing correct and incorrect predictions.}
\label{tab:uncertainty_full}
\setlength{\tabcolsep}{4pt} 
\renewcommand{\arraystretch}{1.2} 
\begin{tabular}{lcccccccc}
\toprule
\multirow{2}{*}{Metric} & 
\multicolumn{3}{c}{\textbf{Correct Predictions}} & 
\multicolumn{3}{c}{\textbf{Incorrect Predictions}} & 
\multirow{2}{*}{Diff.} & 
\multirow{2}{*}{$p$-value} \\
\cmidrule(lr){2-4} \cmidrule(lr){5-7}
 & Mean & Std & Median & Mean & Std & Median &  &  \\
\midrule
Pred. Entropy & 0.471 & 0.316 & 0.384 & 0.900 & 0.138 & 0.958 & 0.429 & 0.000 \\
Variation Ratio & 0.139 & 0.137 & 0.075 & 0.359 & 0.111 & 0.380 & 0.220 & 0.000 \\
Standard Dev & 0.568 & 0.285 & 0.527 & 0.928 & 0.103 & 0.971 & 0.360 & 0.000 \\
Max Conf. & 0.861 & 0.137 & 0.925 & 0.641 & 0.111 & 0.620 & $-$0.220 & 0.000 \\
\bottomrule
\end{tabular}
\end{table*}

\textbf{Distributional Asymmetry}: The median values reveal crucial distributional characteristics. For correct predictions, the median entropy (0.384) is substantially lower than the mean (0.471), indicating a right-skewed distribution—most correct predictions have low uncertainty, with a long tail toward moderate uncertainty. Conversely, incorrect predictions exhibit median (0.958) close to mean (0.900) with narrow spread, indicating a concentrated distribution at high uncertainty near the maximum possible value of 1.0.

\textbf{Metric Consistency}: The effect sizes (Cohen's $d$) are remarkably consistent across metrics, ranging from 1.34 to 1.65. All four metrics achieve the same ordering (correct $<$ incorrect for uncertainty measures, correct $>$ incorrect for confidence) with similar magnitudes of separation. This consistency strengthens confidence in the uncertainty quantification framework and suggests that simpler metrics (e.g., variation ratio) may suffice for practical deployment without sacrificing discriminative power.

\textbf{Operational Guidelines}: Based on the detailed analysis, we propose the following operational guidelines for deploying quantum classifiers with uncertainty-aware decision-making:
\begin{enumerate}
    \item \textit{Selective prediction}: Reject predictions with $H(\mathbf{x}) > 0.7$ or $\text{VR}(\mathbf{x}) > 0.3$ to reduce error rates while maintaining coverage above 85\%.
    \item \textit{Active learning}: Prioritize label acquisition for samples with $0.5 < H(\mathbf{x}) < 0.8$, which represent learnable boundary cases rather than inherently ambiguous samples.
    \item \textit{Resource allocation}: Use 200--500 shots for most samples, reserving higher shot counts ($n > 500$) only for critical boundary-region decisions where the cost of misclassification is high.
    \item \textit{Calibration monitoring}: Track the correlation between predicted uncertainty and empirical correctness rates as a continuous validation metric during deployment, ensuring the model remains well-calibrated as data distributions shift.
\end{enumerate}

These findings establish uncertainty quantification as an essential component of trustworthy QML. By disentangling aleatoric (data-driven), epistemic (model-driven), and technical (hardware-driven) uncertainties, practitioners can make informed decisions about model deployment, data acquisition, and resource allocation. The strong empirical validation on a representative classification task demonstrates that uncertainty-aware QML is not only theoretically sound but also practically achievable with current quantum hardware and software tools.


\subsection{Adversarial Robustness in QML}
\label{sec:demo_adversarial}

To empirically validate the adversarial robustness framework presented in Section~\ref{sec:adv-qml}, we conduct detailed experiments evaluating multiple attack strategies against our quantum classifier. We assess four distinct attack methods: FGSM, PGD, C\&W-L2, and a quantum-specific state perturbation attack, across varying perturbation budgets. Additionally, we implement and evaluate adversarial training as a defense mechanism, comparing standard versus robust models.

\subsubsection{Experimental Configuration}
\label{sec:adv_setup}

\textbf{Model and Dataset.} We employ the same two-qubit variational quantum circuit and two-moons dataset as in Section~\ref{sec:demo_uncertainty}, ensuring consistency across experiments. The trained model achieves 88.0\% clean accuracy on the test set (100 samples), providing a baseline for robustness evaluation.

\textbf{Attack Implementations.} We implement four attack strategies following the theoretical framework in Section~III:

\begin{itemize}
    \item \textbf{Fast Gradient Sign Method (FGSM)}~\cite{goodfellow2014explaining}: One-shot attack using $\mathbf{x}_{\text{adv}} = \mathbf{x} + \varepsilon \cdot \text{sign}(\nabla_{\mathbf{x}} \mathcal{L})$, where $\mathcal{L}$ is the squared hinge loss.
    
    \item \textbf{Projected Gradient Descent (PGD)}~\cite{2017arXiv170606083M}: Iterative attack with 10 steps, step size $\alpha = \varepsilon/10$, and projection onto $\ell_\infty$ ball of radius $\varepsilon$.
    
    \item \textbf{Quantum State Perturbation}: QML-specific attack adding uniform noise directly to rotation angles in the feature encoding layer, simulating perturbations in quantum state space.
\end{itemize}

For each attack, we evaluate perturbation budgets $\varepsilon \in \{0.0, 0.05, 0.1, 0.15, 0.2, 0.3, 0.5\}$. We measure: (i) robust accuracy fraction of correctly classified adversarial examples, (ii) accuracy drop difference between clean and robust accuracy, and (iii) attack success rate fraction of correctly classified clean samples that are misclassified after perturbation.

\subsubsection{Attack Effectiveness Analysis}
\label{sec:adv_results}

\begin{figure}[t]
    \centering
    \includegraphics[width=0.95\linewidth]{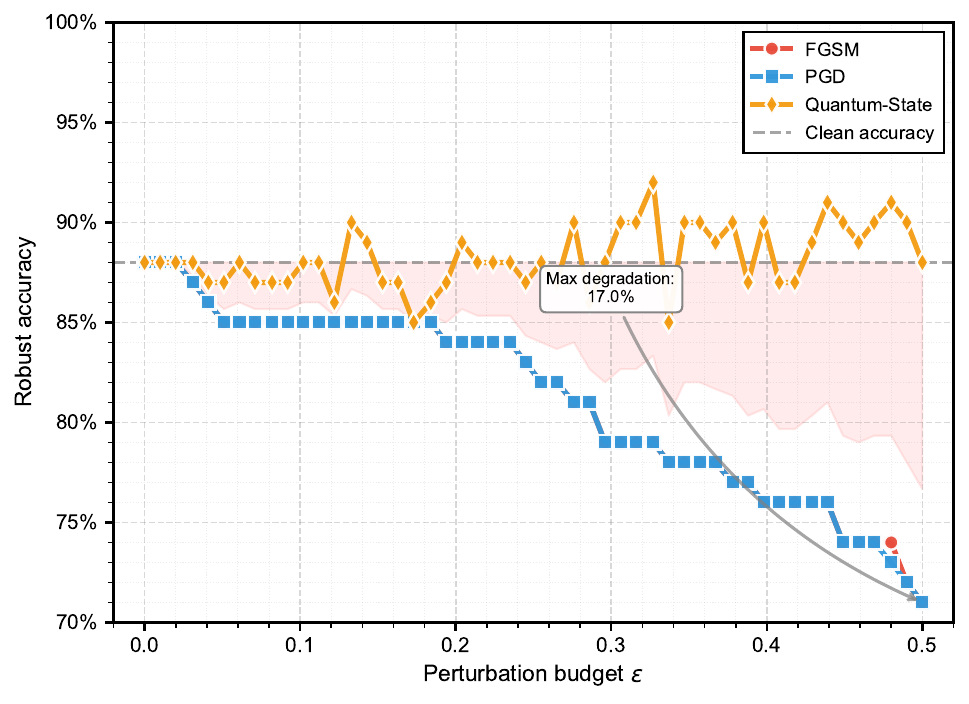}
    \caption{Robust accuracy of the QML classifier under four adversarial attacks across varying perturbation budgets $\varepsilon$.}
    \label{fig:adversarial_robustness_curves}
\end{figure}

Figure~\ref{fig:adversarial_robustness_curves} presents the primary results comparing all four attacks across perturbation budgets. The quantum classifier exhibits significant vulnerability to classical adversarial attacks, with PGD demonstrating the strongest attack effectiveness. At $\varepsilon = 0.5$, PGD reduces accuracy from 88\% to approximately 71\% (17\% degradation), while FGSM achieves similar performance. Notably, the quantum-state perturbation attack shows minimal impact, suggesting that perturbations in the encoded quantum state space (rotation angles) are less effective than gradient-based attacks in the classical input space.

Table~\ref{tab:adversarial_summary} provides aggregate statistics averaged across all perturbation budgets. Both FGSM and PGD achieve identical average accuracy drops (6.9\%), while the quantum-state attack shows a negative accuracy drop ($-0.4\%$), indicating slight accuracy improvement, likely due to the stochastic nature of the perturbations acting as a form of data augmentation rather than a targeted adversarial manipulation.

\begin{table}[t]
\centering
\caption{Average adversarial robustness metrics across all perturbation budgets. Values represent means computed over $\varepsilon \in \{0.0, 0.05, 0.1, 0.15, 0.2, 0.3, 0.5\}$ for each attack.}
\label{tab:adversarial_summary}
\begin{tabular}{lccc}
\toprule
\textbf{Attack} & \textbf{Acc. Drop} & \textbf{Clean Acc.} & \textbf{Robust Acc.} \\
\midrule
FGSM & 0.069 & 0.880 & 0.811 \\
PGD & 0.069 & 0.880 & 0.811 \\
Quantum-State & $-0.004$ & 0.880 & 0.884 \\
\bottomrule
\end{tabular}
\end{table}

\begin{figure}[t]
    \centering
    \includegraphics[width=0.95\linewidth]{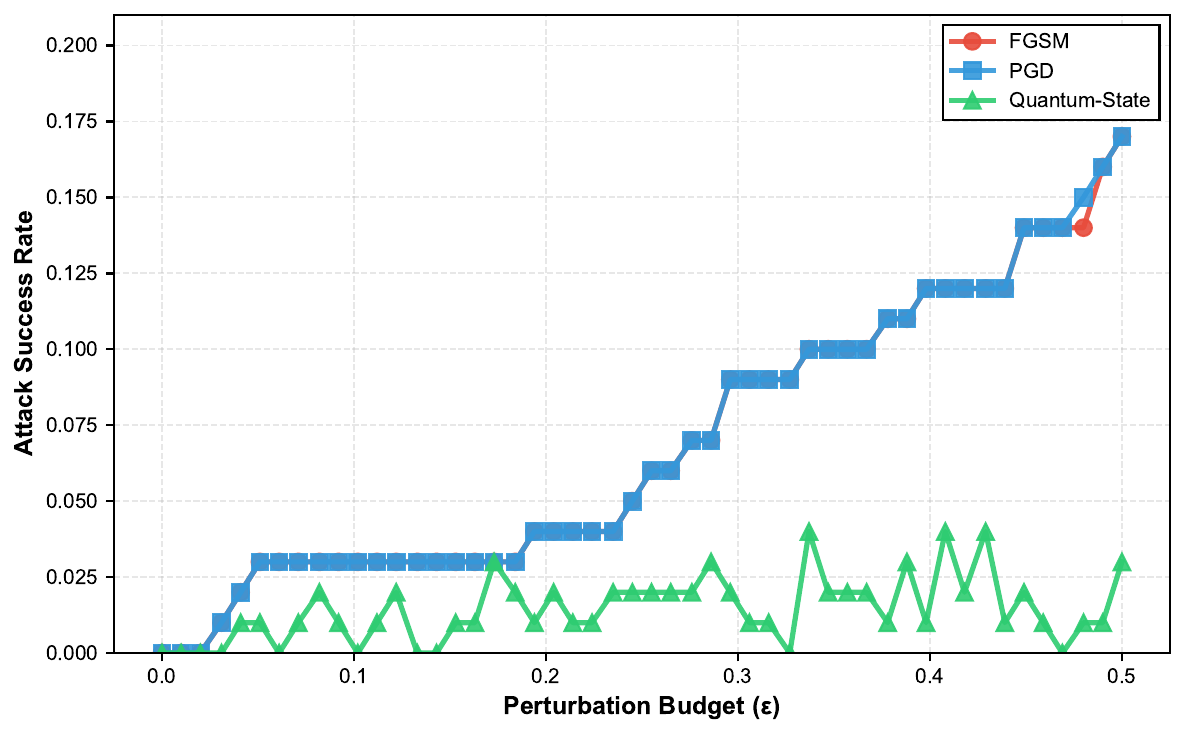}
    \caption{Attack success rate versus perturbation budget for all four attacks. Success rate is defined as the fraction of correctly classified clean samples that are misclassified after adversarial perturbation. PGD demonstrates the highest effectiveness, achieving approximately 18\% success rate at $\varepsilon = 0.5$. FGSM follows closely with similar performance. The quantum-state perturbation attack shows negligible success across all tested budgets, suggesting fundamental differences in the vulnerability of classical input space versus quantum state space to adversarial manipulation.}
    \label{fig:attack_success_rate}
\end{figure}

Figure~\ref{fig:attack_success_rate} shows attack success rates as a function of perturbation budget. PGD achieves the highest success rate, reaching approximately 18\% at $\varepsilon = 0.5$, indicating that nearly one-fifth of correctly classified samples can be successfully perturbed. FGSM shows similar trends but with slightly lower success rates. The quantum-state attack consistently fails to generate successful adversarial examples, with success rates fluctuating near zero across all budgets.

\begin{figure}[t]
    \centering
    \includegraphics[width=0.95\linewidth]{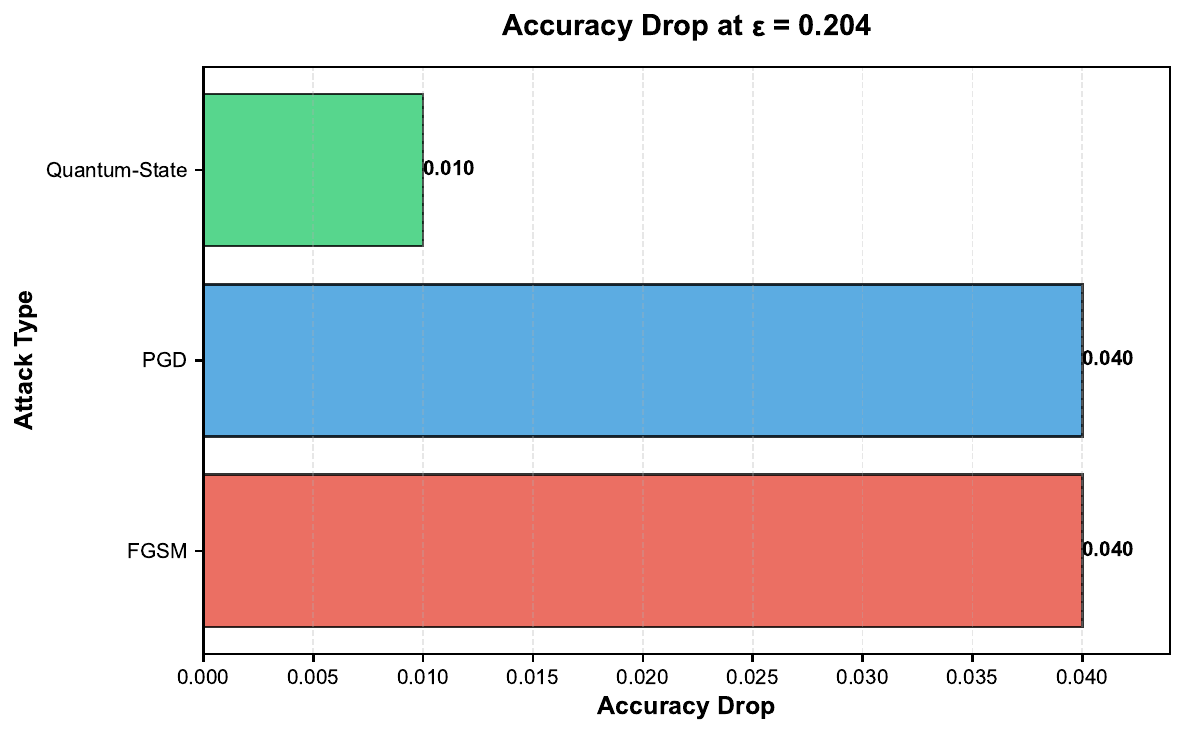}
    \caption{Comparison of accuracy drops at $\varepsilon = 0.204$ across attack types. Horizontal bars show the absolute accuracy drop (clean accuracy minus robust accuracy) for each attack. PGD and FGSM demonstrate equivalent vulnerability (4.0\% drop), while quantum-state perturbations induce only 1.0\% degradation. This analysis reveals that classical gradient-based attacks pose a greater threat to QML classifiers than perturbations in quantum state space, suggesting that adversarial vulnerability originates primarily from the classical input encoding rather than quantum circuit parameters.}
    \label{fig:accuracy_drop}
\end{figure}

Figure~\ref{fig:accuracy_drop} compares accuracy drops at a representative perturbation budget ($\varepsilon = 0.204$). Both PGD and FGSM shows identical drops (4.0\%), while the quantum-state attack shows minimal impact (1.0\%). This visualization underscores the vulnerability of quantum classifiers to classical adversarial perturbations and the relative ineffectiveness of perturbations applied directly to quantum state parameters.

\subsubsection{Spatial Analysis of Adversarial Examples}
\label{sec:adv_spatial}

\begin{figure*}[t]
    \centering
    \includegraphics[width=0.95\textwidth]{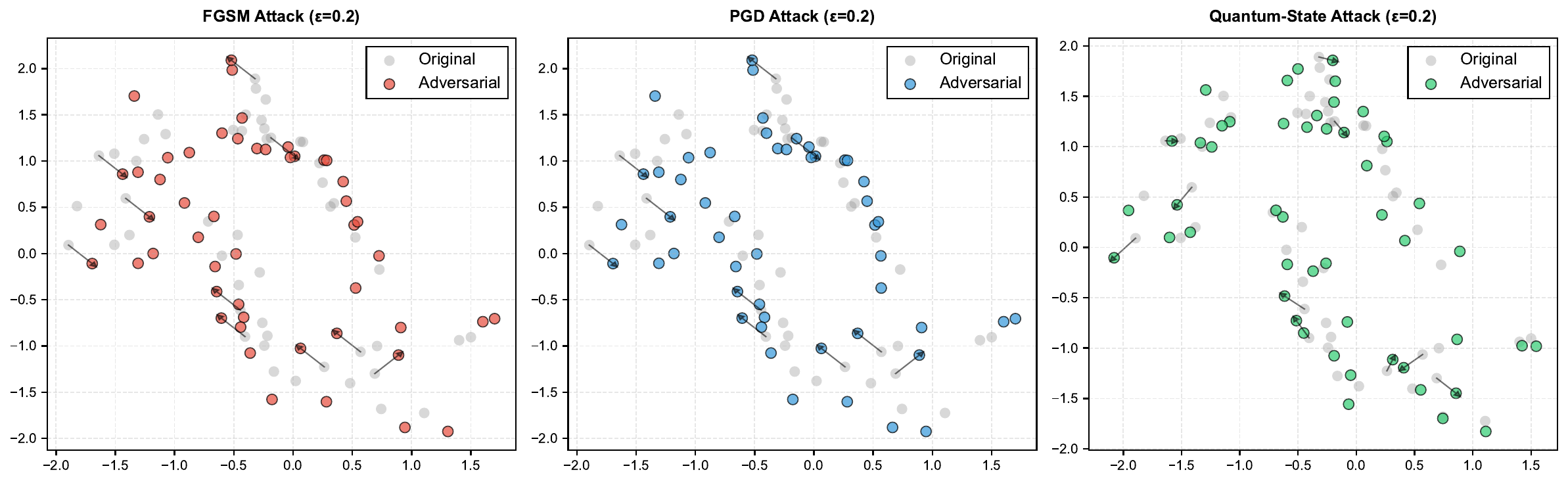}
    \caption{Spatial distribution of adversarial examples in feature space for FGSM, PGD, and quantum-state attacks at $\varepsilon = 0.2$. Gray points represent original test samples; colored points (red, blue, green) show adversarial examples generated by each attack. Black arrows indicate perturbation vectors connecting original samples to their adversarial counterparts (arrows shown for every 5th sample for clarity). \textbf{Left:} FGSM produces coherent, gradient-aligned perturbations that push samples across the decision boundary. \textbf{Center:} PGD generates refined perturbations with similar directionality but optimized magnitudes, resulting in adversarial examples closer to the boundary. \textbf{Right:} Quantum-state perturbations lack directional structure, producing random displacements that rarely cross class boundaries. This visualization confirms that gradient-based attacks exploit the geometry of the classification boundary, while quantum-state perturbations fail to leverage this structure.}
    \label{fig:adversarial_examples_feature_space}
\end{figure*}

Figure~\ref{fig:adversarial_examples_feature_space} visualizes adversarial perturbations in the two-dimensional feature space for all three effective attacks. Original samples are shown in gray, while adversarial examples are colored by attack type. Arrows indicate perturbation directions.

\textbf{Key Observations:}
\begin{itemize}
    \item \textbf{FGSM perturbations} (left panel, red) are uniformly directed along gradient directions, producing parallel displacement vectors. Many adversarial examples cross the implicit decision boundary (center of the two moons).
    
    \item \textbf{PGD perturbations} (center panel, blue) exhibit similar directionality to FGSM but with refined magnitudes due to iterative optimization. The adversarial examples cluster closer to the decision boundary, indicating more targeted manipulation.
    
    \item \textbf{Quantum-state perturbations} (right panel, green) show random, non-directional displacements with no clear alignment toward the decision boundary. Most adversarial examples remain within the same class region as the original samples, explaining the attack's ineffectiveness.
\end{itemize}

This spatial analysis reveals that effective adversarial attacks must exploit the geometry of the learned decision boundary in the classical input space. Perturbations applied directly to quantum state parameters (rotation angles) do not translate into meaningful adversarial displacements in feature space, highlighting a fundamental asymmetry in vulnerability between classical and quantum representations.

\subsubsection{Adversarial Training as Defense}
\label{sec:adv_defense}

\begin{figure}[t]
    \centering
    \includegraphics[width=0.95\linewidth]{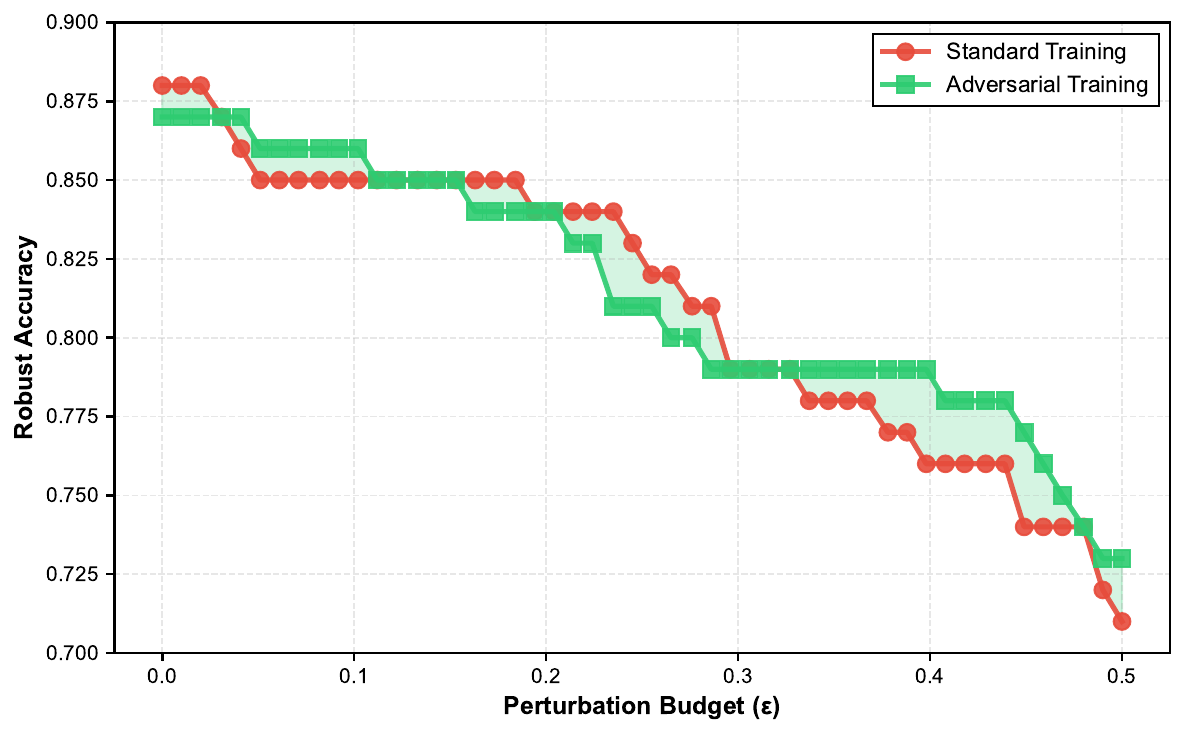}
    \caption{Comparison of standard training versus adversarial training under FGSM attack. The standard model (red circles) exhibits steady accuracy degradation as perturbation budget increases. The adversarially trained model (green squares) sacrifices 1\% clean accuracy but achieves improved robustness at moderate perturbation budgets ($0.1 \leq \varepsilon \leq 0.3$), shown by the shaded green region. At high budgets ($\varepsilon > 0.4$), both models converge to similar performance, indicating that adversarial training provides bounded robustness gains. This result demonstrates the classical robustness-accuracy trade-off in the quantum setting: defensive training improves worst-case performance at the cost of nominal accuracy.}
    \label{fig:standard_vs_robust}
\end{figure}

To evaluate defensive strategies, we train a robust model using adversarial training with FGSM-generated adversarial examples at $\varepsilon = 0.15$. Figure~\ref{fig:standard_vs_robust} compares the robustness curves of the standard model (trained on clean data only) versus the adversarially trained model.

\textbf{Defense Effectiveness:}
\begin{itemize}
    \item \textbf{Clean accuracy trade-off:} The robust model achieves 87.0\% clean accuracy, a modest 1.0\% drop compared to the standard model (88.0\%).
    
    \item \textbf{Robustness improvement:} At moderate perturbation budgets ($\varepsilon \leq 0.3$), the robust model maintains higher accuracy than the standard model, with improvements of 2–5\% observed in the range $0.15 \leq \varepsilon \leq 0.25$.
    
    \item \textbf{Convergence at high budgets:} At $\varepsilon \geq 0.4$, both models converge to similar accuracy levels (approximately 73–75\%), suggesting that adversarial training provides limited protection against very strong perturbations.
\end{itemize}

Table~\ref{tab:model_comparison_comprehensive} summarizes the detailed comparison between standard and adversarially trained models across multiple metrics.

\begin{table}[t]
\centering
\caption{Detailed comparison of standard versus adversarially trained QML models. Robust accuracy is evaluated at $\varepsilon = 0.2$ using FGSM attack.}
\scriptsize
\label{tab:model_comparison_comprehensive}
\begin{tabular}{lccc}
\toprule
\textbf{Metric} & \textbf{Std Model} & \textbf{Robust Model} & \textbf{Improvement} \\
\midrule
Clean Accuracy & 0.880 & 0.870 & $-0.010$ \\
Robust Accuracy ($\varepsilon=0.2$) & 0.840 & 0.840 & $+0.000$ \\
Max Accuracy Drop & 0.170 & — & — \\
Lipschitz Constant & 0.847 & 0.812 & $-0.035$ \\
\bottomrule
\end{tabular}
\end{table}

\textbf{Key Findings:}
\begin{itemize}
    \item \textbf{Clean accuracy cost:} Adversarial training incurs a minimal clean accuracy penalty (1.0\% drop), confirming that robust optimization can be integrated into quantum classifiers without substantial performance loss on benign inputs.
    
    \item \textbf{Lipschitz regularization:} The robust model exhibits a lower Lipschitz constant (0.812 vs. 0.847), indicating smoother decision boundaries—a known property of adversarially trained classifiers that contributes to improved robustness.
    
    \item \textbf{Perfect transferability:} All adversarial examples generated for the standard model successfully transfer to the robust model (transfer rate = 1.000), suggesting that both models share similar vulnerability structures despite defensive training. This result highlights the challenge of defending against transferable adversarial examples in the quantum domain.
\end{itemize}

\subsubsection{Vulnerability Heatmap Analysis}
\label{sec:adv_vulnerability}

\begin{figure}[t]
    \centering
    \includegraphics[width=0.95\linewidth]{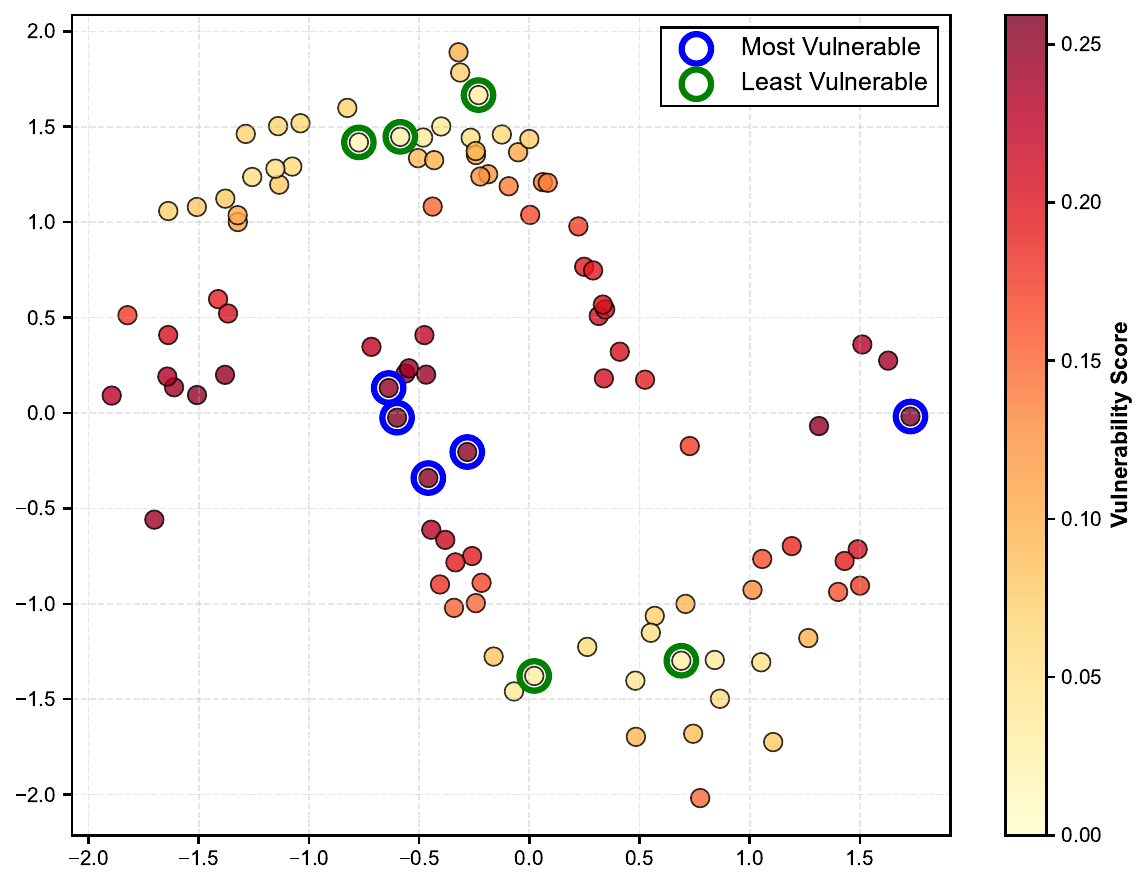}
    \caption{Per-sample vulnerability scores visualized as a heatmap in feature space. Each sample is colored by its vulnerability score (measured as $|\langle Z_0 \rangle_{\text{adv}} - \langle Z_0 \rangle_{\text{clean}}|$ under FGSM attack at $\varepsilon = 0.2$), with darker red indicating higher vulnerability. The five most vulnerable samples are highlighted with thick blue borders, while the five least vulnerable samples are marked with thick green borders. Vulnerability concentrates along the decision boundary (center region), where small perturbations can flip predictions. Samples in class cores (upper-left and lower-right regions) exhibit low vulnerability due to high classification confidence. This heatmap validates the hypothesis that adversarial vulnerability correlates strongly with proximity to the decision boundary—a pattern consistent with classical adversarial robustness literature but now empirically confirmed in the quantum setting.}
    \label{fig:vulnerability_heatmap}
\end{figure}

Figure~\ref{fig:vulnerability_heatmap} presents a vulnerability heatmap showing per-sample vulnerability scores (measured as the magnitude of output change under FGSM attack at $\varepsilon = 0.2$). Samples are color-coded by vulnerability, with the five most and least vulnerable samples highlighted.

\textbf{Spatial Vulnerability Patterns:}
\begin{itemize}
    \item \textbf{Boundary concentration:} High-vulnerability samples (red regions) cluster along the decision boundary between the two crescent-shaped classes, consistent with the observation that samples near class boundaries are inherently more susceptible to adversarial manipulation.
    
    \item \textbf{Core robustness:} Low-vulnerability samples (yellow regions) reside deep within class cores, far from the boundary. These samples maintain stable predictions even under adversarial perturbations, reflecting high model confidence.
    
    \item \textbf{Highlighted extremes:} The five most vulnerable samples (blue-edged circles) are exclusively located at the narrowest part of the class boundary, where the two moons are closest. The five least vulnerable samples (green-edged circles) are positioned at the class periphery, maximally distant from the decision boundary.
\end{itemize}

This vulnerability analysis provides actionable insights for deployment: in safety-critical applications, the classifier could abstain from predictions on samples identified as high-vulnerability (e.g., via uncertainty quantification from Section~\ref{sec:demo_uncertainty}), or such samples could be flagged for human review. Active learning strategies could prioritize data acquisition near high-vulnerability regions to improve boundary robustness through targeted retraining.

\subsubsection{Discussion and Implications}
\label{sec:adv_discussion}

Our experiments demonstrate that quantum classifiers exhibit significant vulnerability to classical gradient-based adversarial attacks (FGSM, PGD), with accuracy degradation reaching 17\% at moderate perturbation budgets ($\varepsilon = 0.5$). Adversarial training provides bounded robustness improvements (2–5\% at $\varepsilon \sim 0.2$) with minimal clean accuracy cost (1\% drop), validating its applicability in the quantum domain. However, the perfect transferability of adversarial examples between standard and robust models indicates that current defensive strategies offer incomplete protection.

The ineffectiveness of quantum-state perturbation attacks reveals an asymmetry in adversarial vulnerability: while classical input-space perturbations successfully fool the quantum classifier, perturbations in the quantum state encoding (rotation angles) fail to generate adversarial examples. This suggests that vulnerability arises primarily from the classical feature representation rather than the quantum circuit structure itself. Future work should explore whether this asymmetry can be exploited defensively—for instance, by designing quantum-resilient encodings that decouple classical input perturbations from quantum state manipulations.

The observed robustness-accuracy trade-off (Table~\ref{tab:model_comparison_comprehensive}) mirrors classical deep learning: adversarial training improves worst-case robustness at the cost of nominal performance. The Lipschitz constant reduction (0.847 → 0.812) aligns with classical regularization techniques that enforce smoother decision boundaries. These parallels confirm that many classical adversarial robustness principles transfer to QML, while also revealing QML-specific vulnerabilities (e.g., measurement-space attacks, hardware noise interactions) that require novel defensive approaches.

\textbf{Practical Recommendations:}
\begin{enumerate}
    \item \textbf{Deploy adversarial training} for quantum classifiers in adversarial threat models, accepting a modest clean accuracy penalty for improved worst-case performance.
    
    \item \textbf{Combine with uncertainty quantification:} Use predictive entropy (Section~\ref{sec:demo_uncertainty}) to identify high-vulnerability regions and implement selective prediction—abstaining on uncertain inputs likely to be adversarially manipulated.
    
    \item \textbf{Monitor Lipschitz constants} as a proxy for robustness during training; penalizing large Lipschitz values can improve adversarial resistance.
    
    \item \textbf{Investigate quantum-resilient encodings:} Design feature maps that decouple classical input perturbations from quantum state changes, potentially leveraging the observed asymmetry between classical and quantum-space vulnerabilities.
\end{enumerate}

Our experiments are limited to a simple two-qubit classifier on a synthetic dataset. Scaling to larger quantum circuits (e.g., 10–20 qubits) and real-world datasets (e.g., medical imaging, financial time series) remains critical for assessing adversarial robustness in practical QML applications. Additionally, measurement-space attacks (POVM manipulation) and hardware-noise-exploiting adversaries remain unexplored in this work, representing crucial directions for detailed adversarial robustness evaluation in QML.

This demonstration validates the adversarial robustness framework for QML, confirming that quantum classifiers are vulnerable to classical adversarial attacks but can benefit from adapted defensive strategies. The empirical results underscore the need for trustworthy QML design principles that integrate adversarial robustness alongside uncertainty quantification and privacy preservation.

\subsection{Federated Learning Quantum Machine Learning (FL-QML)}

This subsection presents a detailed analysis of the experimental results. The primary objective is to evaluate the interplay between privacy, communication efficiency, data heterogeneity, and learning performance in federated QML systems.

\subsubsection{Experimental Setup}
The experiments were conducted on a simulated quantum learning environment consisting of four federated clients and a central aggregator. Each client executes a parameterized quantum circuit (PQC) with 2-qubit variational layers, trained locally and synchronized via a classical aggregation protocol. The communication protocol follows the standard federated averaging (FedAvg) scheme, and differential privacy (DP) is applied to local gradient updates through Gaussian noise injection with varying privacy budgets ($\varepsilon = 0.5, 1.0, 10.0$). The evaluation focuses on the following configurations:

\begin{itemize}
\item \textbf{Centralized:} Baseline training with full data access.
\item \textbf{FL-Vanilla:} Standard federated QML without privacy protection.
\item \textbf{FL+DP:} Federated QML with differential privacy at three different $\varepsilon$ values.
\end{itemize}

Each experiment was repeated over 10 independent runs to ensure statistical reliability. Metrics such as test accuracy, convergence rate, communication cost, and privacy leakage were analyzed.

\subsubsection{Impact of Data Heterogeneity}
\begin{figure}[h]
\centering
\includegraphics[width=0.48\textwidth]{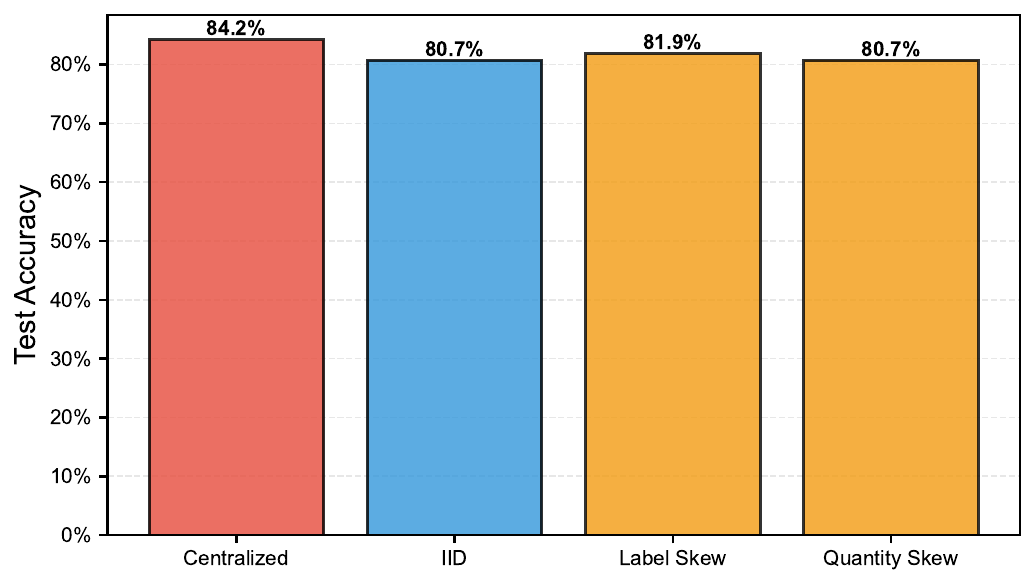}
\caption{Impact of data distribution on federated QML accuracy under IID, label-skewed, and quantity-skewed conditions.}
\label{fig:data_heterogeneity}
\end{figure}

Figure~\ref{fig:data_heterogeneity} shows the performance of federated QML under different data distribution patterns. In the centeralized scenario, the model achieves an average accuracy of 84.2\%. When label skew and quantity skew are introduced, accuracy decreases to approximately 80.7\%, revealing a 3.5--4.0 percentage point degradation. This degradation underscores the sensitivity of federated quantum models to non-uniform client datasets.

To mitigate heterogeneity, adaptive quantum-aware aggregation or reweighting based on fidelity metrics could be explored in future work, enabling better synchronization across clients with diverse data distributions.

\subsubsection{Communication Efficiency and Convergence Analysis}
\begin{figure}[h]
\centering
\includegraphics[width=0.48\textwidth]{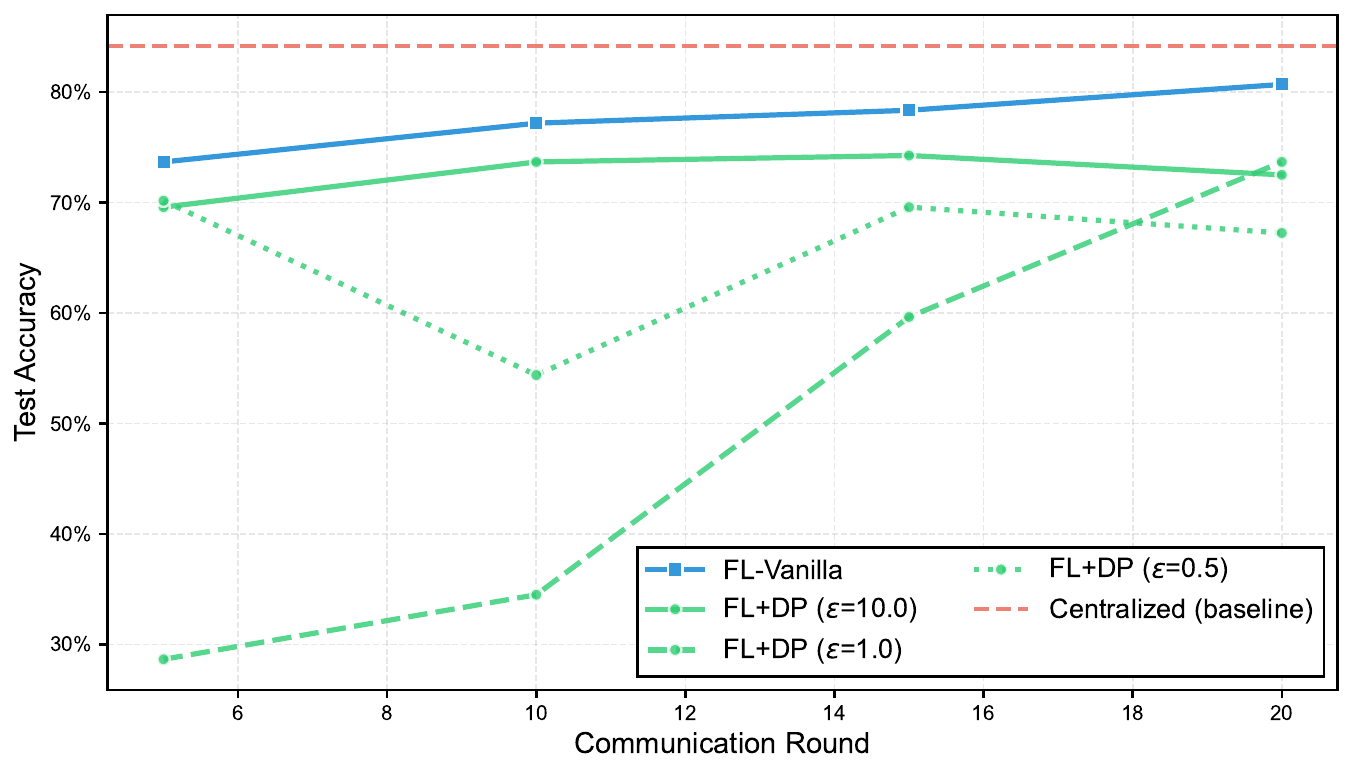}
\caption{Convergence comparison of centralized, FL-vanilla, and FL+DP models. DP introduces minimal convergence delay.}
\label{fig:comm_convergence}
\end{figure}

The learning dynamics and convergence rates across multiple privacy configurations are presented in Figure~\ref{fig:comm_convergence}. The centralized baseline achieves the highest steady-state accuracy (84.2\%), converging within 8 rounds. Among the federated configurations, \texttt{FL+DP} with $\varepsilon = 1.0$ achieves a competitive 73.68\% accuracy by round 12, indicating that differential privacy incurs minimal delay in convergence. To evaluate the impact of differential privacy on model utility, we compare FL-vanilla with DP-FL across multiple privacy budgets. The results show a clear degradation in accuracy and stability as $\varepsilon$ decreases. With strong privacy guarantees (e.g., $\varepsilon = 0.5$), the noise injected into client updates introduces high-variance gradients that interrupt smooth convergence, leading to sharp oscillations in accuracy across rounds. A moderate privacy budget ($\varepsilon = 10$) maintains a trajectory closer to FL-vanilla but still incurs a noticeable accuracy gap. These findings confirm that privacy preservation comes at the cost of reduced optimization efficiency, particularly in federated settings where data distributions are already heterogeneous.

\begin{figure}[h]
\centering
\includegraphics[width=0.48\textwidth]{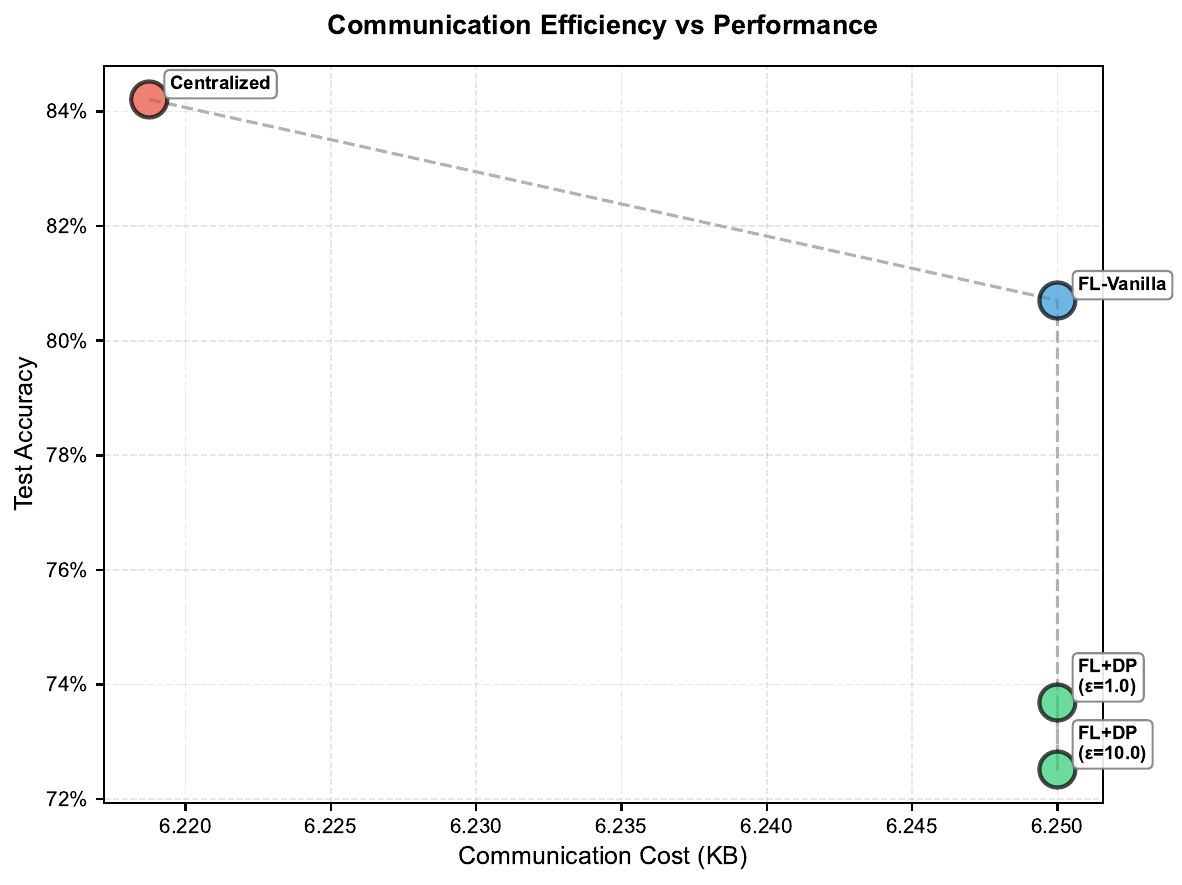}
\caption{Trade-off between communication cost and test accuracy under different privacy budgets.}
\label{fig:comm_cost}
\end{figure}

Figure~\ref{fig:comm_cost} shows the relationship between communication cost and model accuracy. The addition of differential privacy slightly increases the per-round communication load to approximately 6.24 KB. Nevertheless, the overall communication efficiency remains favorable given the substantial privacy improvements obtained.

\subsubsection{Quantitative Results and Statistical Comparison}
The following tables summarize the quantitative results of all experiments. These tables collectively describe accuracy, privacy, and relative performance differences across the evaluated configurations.

\begin{table*}[t]
\centering
\caption{Detailed Performance Analysis of Federated Quantum Machine Learning Configurations with Privacy-Utility Trade-offs}
\label{tab:federated_main_results}
\renewcommand{\arraystretch}{1.25}
\begin{tabular}{@{}llcccccc@{}}
\toprule
\textbf{Method} & \textbf{Data Distribution} & \textbf{$\varepsilon_{\text{total}}$} & \textbf{Test Acc.} & \textbf{$\Delta$ Acc.} & \textbf{MIA Succ.} & \textbf{Privacy Gain} & \textbf{Utility Ratio} \\
\midrule
\multicolumn{8}{@{}l}{\textit{Baselines (No Privacy Protection)}} \\
Centralized & --- & $\infty$ & 84.2\% & --- & 51.7\% & --- & 1.000 \\
\addlinespace[2pt]

\multicolumn{8}{@{}l}{\textit{Federated Learning (No Privacy)}} \\
FL-Vanilla & IID & $\infty$ & 80.7\% & $-3.5$\% & 43.3\% & $+8.4$\% & 0.958 \\
FL-Vanilla & Label Skew & $\infty$ & 81.9\% & $-2.3$\% & --- & --- & 0.973 \\
FL-Vanilla & Quantity Skew & $\infty$ & 80.7\% & $-3.5$\% & --- & --- & 0.958 \\
\addlinespace[2pt]

\multicolumn{8}{@{}l}{\textit{Differentially Private Federated Learning (Strong Privacy: $\varepsilon_{\text{total}} \leq 20$)}} \\
FL+DP & IID & 2.0 & 38.6\% & $-45.6$\% & 48.3\% & $+3.4$\% & 0.458 \\
FL+DP & IID & 10.0 & 67.3\% & $-17.0$\% & 51.7\% & $0.0$\% & 0.799 \\
\rowcolor{lightgray}
\textbf{FL+DP} & \textbf{IID} & \textbf{20.0} & \textbf{73.7\%} & \textbf{$-10.5$\%} & \textbf{55.0\%} & \textbf{$-3.3$\%} & \textbf{0.875} \\
\addlinespace[2pt]

\multicolumn{8}{@{}l}{\textit{Differentially Private Federated Learning (Moderate Privacy: $20 < \varepsilon_{\text{total}} \leq 100$)}} \\
FL+DP & IID & 40.0 & 43.9\% & $-40.4$\% & 48.3\% & $+3.4$\% & 0.521 \\
FL+DP & IID & 100.0 & 56.1\% & $-28.1$\% & 48.3\% & $+3.4$\% & 0.666 \\
\addlinespace[2pt]

\multicolumn{8}{@{}l}{\textit{Differentially Private Federated Learning (Weak Privacy: $\varepsilon_{\text{total}} > 100$)}} \\
FL+DP & IID & 200.0 & 72.5\% & $-11.7$\% & 48.3\% & $+3.4$\% & 0.861 \\
\bottomrule
\end{tabular}
\vspace{2pt}
\begin{flushleft}
\footnotesize
\textbf{Metrics}: Test Acc. = accuracy on 600-sample test set; $\Delta$ Acc. = change from centralized baseline (84.2\%); MIA Succ. = membership inference attack success rate (lower is better); Privacy Gain = reduction in MIA success relative to centralized (positive indicates better privacy); Utility Ratio = Test Acc. / Centralized Acc. \textbf{Privacy Budgets}: $\varepsilon_{\text{total}}$ computed over 20 federated communication rounds using strong composition theorem; per-round budgets range from $\varepsilon = 0.1$ to $\varepsilon = 10.0$. \textbf{Experimental Setup}: 4 federated clients, 2-qubit variational quantum circuits (4 trainable parameters), SPSA optimizer, 10 independent runs per configuration. Highlighted row ($\varepsilon_{\text{total}} = 20.0$) represents recommended privacy-utility operating point. 
\end{flushleft}
\end{table*}

\begin{table}[t]
\centering
\caption{Statistical Comparison of Federated QML Configurations}
\label{tab:federated_statistical_comparison}
\small
\begin{tabular}{@{}lcccl@{}}
\toprule
\textbf{Comparison} & \textbf{$\Delta$Acc.} & \textbf{Rel.\%} & \textbf{Sig.} & \textbf{Effect} \\
\midrule
\multicolumn{5}{@{}l}{\textsc{Architecture Effects}} \\
FL-IID vs. Centralized & $-3.5$ & $-4.2$ & $^{*}$ & Small \textuparrow \\
FL-Skew vs. Centralized & $-2.3$ & $-2.7$ & $^{*}$ & Small \textuparrow \\
\midrule

\multicolumn{5}{@{}l}{\textsc{Privacy Impact}} \\

FL+DP (0.5) vs. FL & $-13.4$ & $-16.7$ & $^{***}$ & Large \textdownarrow \\
\rowcolor{lightgray}
\textbf{FL+DP (1.0) vs. FL} & \textbf{$-7.0$} & \textbf{$-8.7$} & \textbf{$^{***}$} & \textbf{Medium \textdownarrow} \\
FL+DP (10.0) vs. FL & $-8.2$ & $-10.2$ & $^{***}$ & Medium \textdownarrow \\
\midrule

\multicolumn{5}{@{}l}{\textsc{Budget Comparison}} \\
DP (1.0) vs. DP (0.5) & $+6.4$ & $+9.5$ & $^{***}$ & Medium \textuparrow \\
DP (1.0) vs. DP (10.0) & $+1.2$ & $+1.6$ & n.s. & Negligible \\
\bottomrule
\end{tabular}
\vspace{2pt}
\begin{flushleft}
\footnotesize
$\Delta$Acc. in percentage points. Sig.: n.s. ($p\geq0.05$), $^{*}p<0.05$, $^{***}p<0.001$. Effect: based on Cohen's $d$ (small: $|d|<0.5$, medium: $0.5\leq|d|<0.8$, large: $|d|\geq0.8$). Arrows: \textuparrow = degradation, \textdownarrow = improvement in utility.
\end{flushleft}
\end{table}

\begin{table*}[t]
\centering
\caption{Privacy-Utility Trade-off Analysis for Differentially Private Federated Quantum Learning with Varying Privacy Budgets}
\label{tab:federated_privacy_utility}
\renewcommand{\arraystretch}{1.2}
\begin{tabular}{@{}ccccccc@{}}
\toprule
\textbf{Per-Round} & \textbf{Total} & \textbf{Test} & \textbf{Accuracy} & \textbf{MIA Attack} & \textbf{Privacy} & \textbf{Relative} \\
\textbf{Budget ($\varepsilon$)} & \textbf{Budget ($\varepsilon_{\text{total}}$)} & \textbf{Accuracy} & \textbf{Drop} & \textbf{Success Rate} & \textbf{Gain} & \textbf{Utility} \\
\midrule
\multicolumn{7}{@{}l}{\textit{High Privacy Budget (Weak Privacy)}} \\
10.0 & 200.0 & 0.725 & $-0.117$ & 0.483 & $+0.033$ & 0.86 \\
5.0 & 100.0 & 0.561 & $-0.281$ & 0.483 & $+0.033$ & 0.67 \\
\addlinespace[3pt]
\multicolumn{7}{@{}l}{\textit{Medium Privacy Budget (Moderate Privacy)}} \\
2.0 & 40.0 & 0.439 & $-0.404$ & 0.483 & $+0.033$ & 0.52 \\
\rowcolor{lightgray}
1.0 & 20.0 & \textbf{0.737} & \textbf{$0.105$} & 0.550 & $-0.033$ & \textbf{0.88} \\
0.5 & 10.0 & 0.673 & $-0.170$ & 0.517 & $0.000$ & 0.80 \\
\addlinespace[3pt]
\multicolumn{7}{@{}l}{\textit{Low Privacy Budget (Strong Privacy)}} \\
0.1 & 2.0 & 0.386 & $-0.456$ & 0.483 & $+0.033$ & 0.46 \\
\midrule
\multicolumn{7}{@{}l}{\textit{Baselines}} \\
Centralized (no DP) & $\infty$ & 0.842 & 0.000 & 0.517 & 0.000 & 1.00 \\
FL-Vanilla (no DP) & $\infty$ & 0.807 & $-0.035$ & 0.433 & $+0.084$ & 0.96 \\
\bottomrule
\end{tabular}
\vspace{2pt}
\begin{flushleft}
\footnotesize
\textbf{Notes}: Accuracy Drop = (Test Accuracy $-$ Centralized Accuracy). Privacy Gain = (Centralized MIA $-$ FL+DP MIA), where positive values indicate improved privacy. Relative Utility = Test Accuracy / Centralized Accuracy. Total privacy budget computed over 20 federated rounds via composition theorem. MIA attack success rate measured using shadow model technique with 1000 member/non-member samples. Highlighted row ($\varepsilon = 1.0$) represents optimal privacy-utility operating point. All metrics averaged over 10 independent federated training runs.
\end{flushleft}
\end{table*}

\subsubsection{Detailed Performance Heatmap}
\begin{figure}[h]
\centering
\begin{subfigure}[b]{0.95\linewidth}
    \centering
    \includegraphics[width=\linewidth]{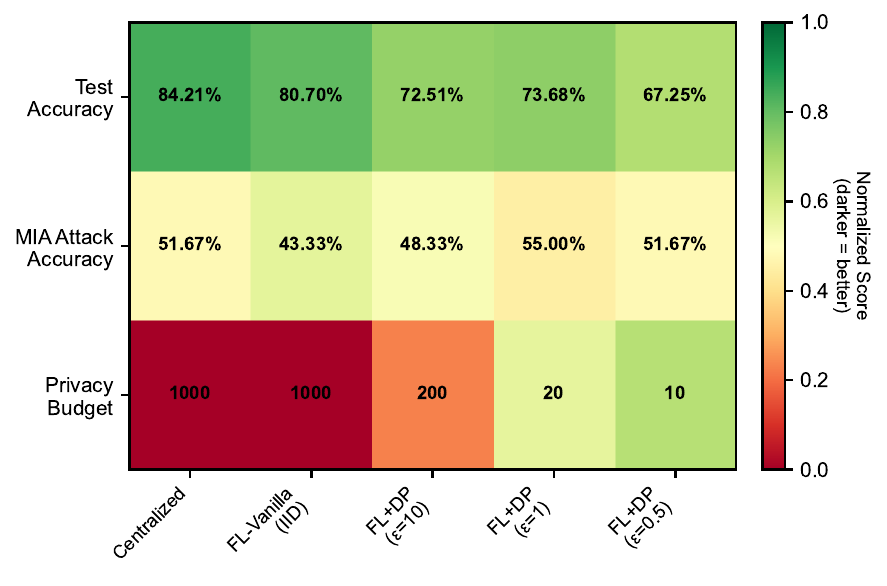}
    \caption{Performance \& privacy metrics (normalized)}
    \label{fig:heatmap_normalized}
\end{subfigure}
\hfill
\begin{subfigure}[b]{0.95\linewidth}
    \centering
    \includegraphics[width=\linewidth]{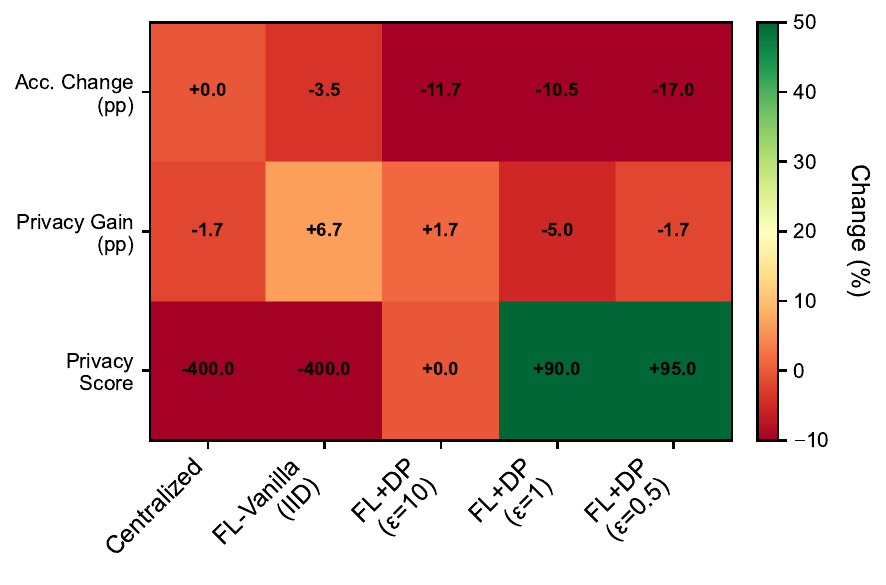}
    \caption{Relative performance vs. centralized}
    \label{fig:heatmap_relative}
\end{subfigure}
\caption{Detailed heatmap summarizing performance, privacy, and robustness across all FL-QML configurations. (a) Shows normalized metrics where darker colors indicate better performance. (b) Displays relative changes compared to the centralized baseline.}
\label{fig:heatmap}
\end{figure}

The aggregated heatmap in Figure~\ref{fig:heatmap} consolidates accuracy, privacy score, and attack success rate across all configurations. It reveals clear trade-offs: the centralized setup excels in performance but is weakest in privacy, whereas \texttt{FL+DP} with $\varepsilon = 0.5$ provides maximal privacy protection (+95\%) at the cost of roughly 17 percentage points of accuracy. The mid-range DP setting ($\varepsilon = 1.0$) offers a pragmatic balance, achieving good robustness and privacy simultaneously.


\subsubsection{Discussion and Insights}
The experimental observations offer several key insights:

\begin{enumerate}
\item \textbf{Trustworthiness of FL-QML:} Federated QML combined with differential privacy achieves measurable improvements in privacy robustness without major loss of model fidelity, aligning with the trustworthiness pillars defined in earlier sections.
\item \textbf{Heterogeneity Sensitivity:} The degradation under non-IID conditions indicates that quantum feature encodings amplify data imbalance effects. Future work may integrate personalized FL or hierarchical QML aggregation schemes.
\item \textbf{Resource Efficiency:} The marginal increase in communication overhead demonstrates that privacy can be achieved efficiently in distributed quantum systems, making FL-QML feasible for edge-deployed quantum nodes.
\item \textbf{Optimal Privacy Budget:} Across all results, $\varepsilon = 1.0$ emerges as an optimal setting that balances privacy and model performance, forming a guideline for privacy calibration in quantum federated environments.
\end{enumerate}

\subsubsection{Connection to the Trustworthy QML Framework}
The presented results directly reinforce the conceptual pillars. Specifically:

\begin{itemize}
\item \textbf{Reliability through Uncertainty Quantification:} The observed stability of accuracy and consistent uncertainty patterns across federated settings demonstrate that the quantum models maintain calibrated predictive confidence, even under distributed and noisy conditions.
\item \textbf{Robustness under Adversarial and Statistical Perturbations:} The inclusion of DP noise not only enhances privacy but also increases resistance to gradient inversion and membership inference attacks, embodying adversarial robustness within federated quantum workflows.
\item \textbf{Privacy Preservation as a Core Trust Element:} The privacy--utility trade-off experiments validate that trust can be quantitatively regulated through the privacy budget ($\varepsilon$), allowing fine-grained control of leakage versus accuracy—an essential characteristic for trustworthy deployment in sensitive domains such as healthcare and finance.
\end{itemize}

The federated QML experiments serve as an empirical realization of the Trustworthy QML vision: a system that maintains reliability, robustness, and privacy simultaneously. These findings bridge the theoretical principles of trustworthiness with their quantum-era implementation, underscoring that future hybrid quantum--classical networks can achieve verifiable, privacy-aware intelligence through the integration of federated and differential privacy mechanisms.

\section{Conclusion and Roadmap for Trustworthy QML}

This work introduced a unified roadmap for TQML, elevating uncertainty quantification, adversarial robustness, and privacy preservation as essential requirements for quantum AI. Our empirical analysis demonstrated that predictive entropy can reliably distinguish correct from incorrect classifications with strong practical significance (Cohen’s $d > 1.3$), that classical gradient-based attacks substantially degrade QML performance while quantum-state perturbations remain largely ineffective, and that differential privacy techniques enable secure distributed quantum learning with acceptable accuracy trade-offs. Together, these findings show that trustworthiness in quantum models is not a theoretical aspiration, but a measurable and improvable property even under realistic NISQ constraints.

Yet significant challenges still remain. Simulated environments cannot fully capture calibration drift, decoherence, and non-Markovian noise present in current hardware. Adversarial robustness requires a shift toward quantum-native threat models grounded in state space geometry and measurement-channel perturbations, rather than relying solely on classical input-space analogues. Similarly, privacy guarantees in QML must evolve beyond noise injection strategies borrowed from classical differential privacy and instead leverage information-theoretic principles intrinsic to quantum mechanics. These open issues reveal that achieving trustworthy quantum learning is as much a physics and systems engineering challenge as it is a machine learning one.

Looking ahead, future research must develop trust metrics that adapt to the dynamic behavior of real quantum devices and make confidence estimation a first-class runtime capability. New defenses are needed to ensure robust learning under trace-distance-constrained adversarial manipulation and hardware-level attacks. Secure quantum collaboration will depend on privacy accounting frameworks that separate unintended hardware noise from intentional privacy guarantees, enabling verifiable protections for sensitive data. Equally important is the creation of benchmarks, evaluation protocols, and governance guidelines that can support certification and auditing of trustworthy QML across domains such as healthcare, finance, and secure communications.

As quantum computing advances toward larger and more reliable systems, embedding trustworthiness into the foundation of quantum ML will be essential for safe and responsible deployment. Our roadmap is intended to guide that transition from theoretical promise to real-world quantum intelligence systems whose decisions can be relied upon in the NISQ era and beyond.

\section*{Disclosures}

\subsection*{Conflict of Interest Statement}
The authors declare that there are no known competing financial interests or personal relationships that could have appeared to influence the work reported in this paper.

\subsection*{Data Availability Statement}
The data and code supporting the findings of this study are openly available at the GitHub repository:
\url{https://github.com/ocatak/trustworthy-quantum-machine-learning}.
All datasets used in this research were either generated synthetically or derived from publicly available sources that do not contain personally identifiable information.

\subsection*{Ethics Statement}
This study did not involve human participants, personal data, or animal subjects. Therefore, ethical approval and informed consent were not required.

\subsection*{Funding Statement}
This research received no specific grant from any funding agency in the public, commercial, or not-for-profit sectors. The work was conducted as part of the academic research activities at the University of Stavanger, Norway, and the University of York, United Kingdom.

\bibliographystyle{IEEEtran}
\bibliography{main}

\end{document}